\documentclass{jpp}
\usepackage{graphicx}
\usepackage{float}
\usepackage{subcaption}
\usepackage{url}

\usepackage[utf8]{inputenc}
\usepackage[T1]{fontenc}
\usepackage{amsmath}

\title{Phase-Space Non-Integrability of Alpha Particles in Near-Omnigeneous Stellarators}

\author{A. L. Lachmann \aff{1}
  \corresp{\email{a.lachmann@columbia.edu}},
  A. R. Knyazev \aff{1}
 \and E. J. Paul \aff{1}}

\affiliation{\aff{1}Department of Applied Physics and Applied Mathematics, Columbia University, New York, New York 10027, USA }
\begin{document}

\maketitle

\begin{abstract}
In a burning plasma, fusion-born alpha particles sustain the fusion reaction by heating the plasma core. Alpha-particle confinement in stellarators is sensitive to both deviations from omnigeneity and time-dependent Alfvénic perturbations. 
%Overlapping phase space islands induced by Alfvén waves can drive EP transport and induce stochastic regions in phase space. 
Traditional Poincaré maps rely on symmetry assumptions or dimensionality reductions that fail for stellarators with quasisymmetry (QS) error or multi-harmonic Alfvénic perturbations. To address these challenges, we investigate weighted Birkhoff averaging (WBA) as a diagnostic to detect the onset of chaos in guiding-centre trajectories subject to general magnetic perturbations. 
WBA exploits the super-convergence of time averages along quasiperiodic orbits, discriminating regular from chaotic trajectories using any smooth observable rather than an exact invariant of motion.
Applying WBA to the canonical momentum $P_\eta$, we show that convergence tracks underlying orbit regularity even when quasisymmetry is imperfect and $P_\eta$ is not strictly conserved. 
Convergence is measured by the digit accuracy of the weighted average along guiding-centre orbits in fields with imperfect omnigeneity, with and without Alfvén eigenmodes computed using AE3D \citep{spongstability}. We use this diagnostic to map phase-space chaos as a function of equilibrium invariants to identify transport barriers in unperturbed and perturbed stellarator equilibria. 
%This diagnostic can quantify phase space chaos in the absence of symmetry, making it applicable to study EP phase space driven behaviors. 
This work demonstrates that chaos induced by many Alfvén harmonics is often not apparent from the largest harmonic alone. We also find evidence that the onset of chaotic behaviour is sensitive to the local radial profile of the wave and not only to its amplitude at the resonance.
\end{abstract}

\section{Introduction}
The deuterium–tritium fusion reaction produces 3.5 MeV alpha particles, which constitute an energetic particle (EP) population relative to the plasma bulk and play a central role in sustaining a hot plasma core in magnetically confined fusion devices. Beyond the fusion reaction, heating mechanisms such as Ion Cyclotron Radio-Frequency (ICRF) and Neutral Beam Injection (NBI) can drive EP populations. Future experiments and fusion pilot plants will rely on the confinement of EP species for heating, including SPARC \citep{SPARC_EPs, SPARC_Physics}, Infinity Two \citep{Infinity2EPs}, and ITER \citep{ITER}. 
%SPARC relies on ICRF and fusion born fast ions as necessary heating to maintain a burning plasma \citep{SPARC_EPs, SPARC_Physics}. 
This work focuses on fusion-born alpha particles in reactor-scale optimised stellarators. 

Recent stellarator optimisation has achieved excellent guiding-centre alpha confinement in the absence of waves \citep{quasisym_landreman_paul,goodsquid}. Alpha particles may resonate with small deviations from perfect omnigeneity, producing phase-space islands even in well-optimised fields \citep{chambliss_fast_2025,Foster}. Isolated island chains in phase space situated away from the plasma edge generally do not impact EP losses to the reactor wall. However, the overlap of such islands may be a primary driver of transport. As EPs are well confined in modern stellarators \citep{quasisym_landreman_paul, goodsquid}, time-dependent magnetic perturbations, such as Alfvén eigenmodes, may become one of the primary mechanisms for EP transport \citep{Paul_Mynick_Bhattacharjee_2023}. 

Alfvén eigenmodes (AEs) are MHD waves commonly observed in both stellarators and tokamaks and are widely regarded as a source of EP transport \citep{transportstellvtokamak,heidbrink2008basic}. Shear Alfvén waves (SAWs) are magnetic perturbations that propagate parallel to the equilibrium magnetic field, deforming the magnetic field in the transverse direction. In stellarators, passing alphas can resonate with SAWs when a linear combination of their poloidal and toroidal transit frequencies matches the mode frequency. These resonances create islands in phase space that may overlap with pre-existing islands caused by imperfections in omnigeneity. 
Sideband resonances arise from coupling through the quasisymmetric periodicity of a stellarator, leading to the excitation of islands adjacent to, but distinct from, a primary resonance \citep{transportstellvtokamak,Paul_Mynick_Bhattacharjee_2023}.
Overlap between sideband islands, or with islands arising from deviations from perfect omnigeneity, can produce chaotic regions that enhance transport.  Quantifying phase-space chaos is therefore essential for characterising SAW-driven transport. 
%IF DIFFUSION: In this work, we assess the relative importance of diffusion driven by omnigeneity (or quasi-symmetry) error compared to pitch-angle scattering, with the goal of quantifying how QS imperfections interact with Alfvénic perturbations to influence energetic-particle transport.

Assessing the impact of quasisymmetric (QS) imperfections requires a detailed understanding of chaotic regions in EP phase space. Existing methods to measure chaos include converse Kolmogorov–Arnold–Moser (converse KAM) theory, which measures tangent vector rotation along trajectories in phase space to distinguish between intact and broken invariant tori \citep{rotationmetric}. However, converse KAM is limited in that it only proves nonexistence of tori, and therefore it cannot distinguish between phase-space islands and chaos. In QS stellarators and tokamaks, intact invariant tori, islands, and chaotic regions are typically visualised with a Poincaré section, where the known background symmetry reduces the orbit dynamics to two dimensions \citep{White_2011, Paul_Mynick_Bhattacharjee_2023}. The applicability of Poincaré sections is largely limited to systems with symmetries that permit dimensionality-reducing assumptions \citep{Paul_Mynick_Bhattacharjee_2023}. 
%Passing and trapped particles exhibit different dynamics, and therefore different mapping techniques are necessary for dimensionality reduction (\cite{chambliss_fast_2025}). 
%In the absence of an AE or QS breaking perturbation, both the particle energy and the canonical momentum are invariants of motion, which simplifies the dimensionality reduction. 
For example, a 2D Poincaré section can be defined in the presence of a single SAW harmonic in perfect quasisymmetry through an invariant effective energy. 
%This effective invariant arises from a transformation of the equilibrium invariants into the phase of the harmonic. 
With multiple symmetry-breaking perturbations carrying distinct phases, or in an imperfectly quasisymmetric background, such a 2D Poincar\'{e} section is not defined. This motivates alternative methods to classify chaos induced by island overlap. 

We employ weighted Birkhoff averaging (WBA) \citep{meissdun} as an alternative to Poincaré sections to determine whether a trajectory is chaotic. WBA leverages the property that integrable Hamiltonian flows are periodic or quasiperiodic. The WBA is superconvergent for integrable flows, with faster than $T^{-h}$ convergence for any positive integer $h$ \citep{meissdun}, where $T$ is the length of the trajectory segment over which the average is taken. Conversely, in chaotic regimes, convergence is slow and irregular, approximately scaling as $T^{-\frac{1}{2}}$. 
When phase-space islands overlap, chaotic motion is generically expected \citep{chirikov1979} and trajectories vary irregularly in their sampling of phase space, indicating the WBA will not converge superpolynomially. 
%For regular trajectories, the WBA still converges much faster than for chaotic trajectories even when no integral of motion is available, as demonstrated in \cite{meissdun}. 
The convergence rate degrades to polynomial at and around low-order resonances \citep{meissdun}. WBA quantifies departures from periodic and quasiperiodic motion, and can be used to identify transport barriers, or regions of phase space with intact particle trajectories. 

Previous work by \cite{MRuth} employs Birkhoff reduced-rank extrapolation (RRE), which measures convergence of ergodic averages with dynamic weights. Trajectories are classified by the RRE residual, which measures how well a short, adaptive linear filter can make an observable time invariant, permitting an adaptive number of map iterations. \cite{MRuth} apply this to both the standard map and a two-dimensional magnetic field line map obtained over a stellarator field period without time-dependent perturbation. 
% The standard map and the stellarator field period map required different chaos classification thresholds for convergence, with a less stringent criterion applied to the noisier stellarator map. 
In this work, WBA is instead applied to energetic particle guiding-centre phase space, where symmetry breaking and multiple time-dependent harmonics preclude a two-dimensional map.

%Quantifying phase space chaos is necessary to better characterize underlying wave-particle interactions, as their behavior has direct impacts on MHD instability onset and nonlinear evolution. Alpha particles can exchange energy with Alfvén modes, driving or damping the wave depending on the local EP distribution function gradient. 
%The direction of energy transfer between the EP population and the wave is determined by the distribution function gradient. 
%Strong alpha distribution gradients are implicit in the fusion birth distribution, or form through ion resonance RF or NBI heating. These alpha-driven modes increase EP transport though the creation of phase space structures that convect in phase space \citep{Particles_in_chirping}. Chaotic transport induced by both AEs and omnigeneity error can lead to nonlinear saturation. 
%WBA based maps can provide a helpful tool for understanding alpha particle resonant behavior. Understanding alpha particle coherence with modes is useful for modeling AE saturation, which depends on alpha particle diffusivity and transport through the resonance. 

We address how interactions between SAWs and deviations from omnigeneity affect energetic alpha transport by identifying regions of chaos and transport barriers. We validate WBA against Poincaré maps and calibrate a chaos threshold. We measure alpha-particle transport across the entirety of phase space, even when sufficient symmetries to make a Poincaré map do not exist.
% as most optimized stellarators are not perfectly omnigeneous due to engineering and optimization constraints. 
We study the effect that imperfections in omnigeneity have on transport near the SAW particle resonance. Additionally, we demonstrate that the confluence of both omnigeneity error and multi-harmonic overlap induces transport in ways that single-resonance analysis misses. We integrate the guiding-centre equations of motion with AEs calculated in AE3D \citep{spongstability} and apply weighted Birkhoff averaging (WBA) to the canonical momentum to achieve these goals. 

Applied 3D fields have been shown to suppress AEs in tokamaks such as ASDEX Upgrade \citep{ASDEX} by redistributing fast ions and creating chaotic regions in phase space. 
Previous work \citep{NSTX} has developed the concept of ``phase-space engineering,'' or deliberately applying 3D magnetic perturbations to evacuate EP phase space around the resonance, and thereby suppress the AE drive. With a more targeted approach, background field perturbations could be designed to resonate near the main AE resonance, creating a thin chaotic layer that flattens the local distribution function while maintaining overall EP confinement. 
% locally modify the distribution function around AE resonances. This differs from the approach of \citet{NSTX}, as one could strategically position resonances and transport barriers while preserving overall EP confinement. 
% WBA could be used to intentionally craft chaotic regions to prevent phase correlation to the wave. 
WBA would be a useful tool for quantifying chaos in the desired regions for phase-space tailoring, and to confirm the persistence of the invariant tori that bound these regions. 
% Maintaining invariant tori around the targeted regions of chaos is important, as they may act as transport barriers preventing alpha particle loss. 

%Prior to recent work by \cite{Bindel_2023}, performing guiding-center tracing during stellarator optimization was considered too expensive. 
Recent work \citep{Bindel_2023,landreman2026bayesian} has demonstrated the use of guiding-centre tracing to directly optimise stellarators with respect to their alpha-particle confinement, as opposed to optimising with respect to proxies for confinement. Guiding-centre trajectory-based chaos diagnostics such as WBA may be used to understand and diagnose any existing transport barriers that result from alpha-confinement-based optimisation. Alternatively, WBA could be directly applied during the optimisation loop, to optimise deliberately for transport barriers, or for chaotic regions at target locations in phase space. 

%In a similar vein, guiding-center trajectory based chaos diagnostics like WBA may be directly applied during the optimization loop, to optimize magnetic fields for `phase space engineering.'

In Section \ref{sec:lagrangian}, we discuss the guiding-centre equations of motion in the presence of a SAW, and motivate the choice of observable for the weighted Birkhoff average. Section \ref{sec:poincare} introduces the effective invariant of guiding-centre motion with a single-harmonic SAW, and uses this to define the kinetic Poincaré map for trapped and passing particles with a single-harmonic SAW in an exactly quasisymmetric background. The weighted Birkhoff average is introduced in depth in Section \ref{sec:WBA}, along with the motivations for the chosen convergence parameters. Section \ref{sec:equilibrium_chaos} uses WBA to investigate stochasticity arising from QS error across all phase space, including barely trapped particles, which cannot be visualised with Poincaré maps.  Section \ref{sec:qs_error} introduces single-harmonic SAWs to QS stellarators, investigating the relationship between the WBA phase-space map and the corresponding Poincaré maps. We note the effects of background QS error on the onset of chaos due to a SAW. Section \ref{sec:multiharmonic} introduces multi-harmonic AEs to the QA equilibrium and maps quasi-isodynamic (QI) field phase space. 
\section{Perturbed guiding-centre Lagrangian}
\label{sec:lagrangian}

To integrate alpha-particle dynamics in perturbed stellarator fields, we employ the guiding-centre Lagrangian formalism. The Lagrangian \citep{littlejohn_variational_1983}
$$
\mathcal{L}\left(\boldsymbol{R}, \dot{\boldsymbol{R}}, v_{\|}\right)=q\left(\boldsymbol{A}_0+\alpha \boldsymbol{B}_0+\frac{M_i v_{\|}}{q B_0} \boldsymbol{B}_0\right) \cdot \dot{\boldsymbol{R}}-\frac{M_i v_{\|}^2}{2}-\mu B_0-q \delta \Phi
$$
 describes the evolution of the guiding-centre position $\boldsymbol{R}$ and velocity $\dot{\boldsymbol{R}}$ according to interactions with the equilibrium vector potential $\boldsymbol{A}_0$, the equilibrium magnetic field $\boldsymbol{B}_0$, the scalar potential $\delta \Phi$, and the shear Alfvén perturbation given by $\delta\boldsymbol{B}=\nabla\times\alpha\boldsymbol{B}_0$. The magnetic moment, $\mu$, is an adiabatic invariant, defined as the ratio of the perpendicular energy to the magnetic field strength.
 % at the location of the particle.
 The parallel velocity, $v_{\|}$, is the component of the velocity along $\boldsymbol{B}_0$. $M_i$ is the alpha-particle mass. Without a SAW, the total energy, $E = E_{\|} + \mu B$, where $E_{\|} = M_i v_{\|}^2/2$, is set to 3.5 MeV. 
 % This energy corresponds to $E = E_{\|} + \mu B(s, \theta, \zeta)$. 
 In equilibrium, or the unperturbed system, we assume $\delta \Phi=0$.

For marker particle tracing, we employ FIRM3D, a stellarator modelling code which integrates the collisionless Littlejohn Lagrangian equations of motion \citep{FIRM3DJOSS}. FIRM3D supports both unperturbed and perturbed dynamics, evolving particle trajectories in magnetic fields with or without Alfvénic perturbations.
The particles are traced in Boozer coordinates $(\psi, \theta, \zeta)$, where $2\pi \psi$ is the toroidal flux, $2\pi \psi_P$ is the poloidal flux, $\theta$ is the poloidal angle, and $\zeta$ is the toroidal angle \citep{Boozer1983}. We use the normalised toroidal flux $s = \psi/\psi_b$ as a radial coordinate, where $2\pi \psi_b$ is the toroidal flux at the plasma boundary. In these coordinates, the magnetic field is 
$$
\begin{gathered}
\boldsymbol{B}_0=\nabla \psi \times \nabla \theta-\iota(\psi) \nabla \psi \times \nabla \zeta, \\
\boldsymbol{B}_0=G(\psi) \nabla \zeta+I(\psi) \nabla \theta+K(\psi, \theta, \zeta) \nabla \psi,
\end{gathered}
$$
 in the contravariant and covariant forms, respectively. $\iota(\psi)$ is the rotational transform, $G(\psi)$ is the poloidal current function, $I(\psi)$ is the toroidal current function, and $K$ is the radial covariant component.
 % related to $p^\prime(\psi)$.
Since $K(\psi, \theta, \zeta)$ scales with $\beta$, in the following particle tracing we enforce $K = 0$; see \cite{alex} for further discussion.

It is useful to describe the particle motion in Boozer coordinates to make quasisymmetry apparent.  
% By the definition of quasisymmetry, we can construct a linear combination of Boozer coordinates such that the magnetic field strength becomes symmetric.  
In this way, we define a general coordinate system that is applicable across all choices of quasisymmetry, differing only by two constants that specify the helicities. These symmetry coordinates are defined as $\chi=M \theta -N \zeta \text { and } \eta=M^{\prime} \theta-N^{\prime} \zeta$, where $M$ and $N$ are the poloidal and toroidal helicities, respectively. $M^\prime$ and $N^\prime$ are chosen in accordance with symmetry to enforce an equilibrium field strength independent of $\eta$. 
%Additionally, $M^\prime$ and $N^\prime$ are chosen such that $M^{\prime} N \neq M N^{\prime}$, ensuring a valid Jacobian. 
Thus, $\eta$ is an ignorable coordinate and the corresponding canonical momentum, $P_\eta$, is conserved in perfect quasisymmetry. We can define symmetries and their corresponding combinations of Boozer coordinates: $M=1, N=0$ for quasiaxisymmetry; $M=1$ and $N = \pm N_{P}$ for quasihelical symmetry, where $N_{P}$ is the number of field periods; and $M=0, N=\pm N_P$ for quasipoloidal symmetry. We now define the canonical momentum for any helicity:
\begin{equation}
P_\eta=\frac{\partial \mathcal{L}}{\partial \dot\eta}= \frac{q \left( M \psi_P - N \psi \right)}{ M N^\prime - M^\prime N } - \frac{q \left( NI(\psi) - M G(\psi)\right)}{\left( M N^\prime - M^\prime N \right)} \left(\alpha + \frac{M_i v_\|}{q B_0} \right).
\label{eq:peta}
\end{equation}

$P_\eta$ is conserved in the perfectly quasisymmetric equilibrium without a SAW, when $\alpha = 0$. 
% Thus, $P_\eta$ is a strong candidate for isolating the effects of deviations on the particle motion. 
We choose $P_\eta$ as our smooth observable to average, as detailed in Section \ref{sec:WBA}.  In stellarators close to quasisymmetry, $P_\eta$ varies quasiperiodically, so its weighted average should converge rapidly. The averaging of $P_\eta$ is still applicable even in cases when it is not a perfect invariant. Since the WBA damps the endpoint contributions to the orbit average, when a smooth observable oscillates quasiperiodically along a regular orbit, the residual of the weighted average decays faster than any power of $T$. 

In quasi-isodynamic equilibria, we calculate $P_\eta$ using the helicity of the omnigeneity, $M=0$, $N=N_P$, although no continuous symmetry exists in this case. We observe rapid convergence of the WBA of $P_\eta$ even where $P_\eta$ is not an exact invariant. 
% This is consistent with the WBA depending on the quasiperiodicity of the observable rather than on its exact invariance. 
This reflects that regular trajectories may persist even in the absence of a global invariant, and that the WBA converges rapidly for trajectories on invariant tori because the oscillations of $P_\eta$ along such orbits are quasiperiodic. 
%Additionally, $P_\eta$ maintains a strong dependence on $\psi$, lending to physically interpretable results indicating when particles sample spaces of
 % Since WBA converges for smooth periodic functions, it identifies chaos even when far from QS and $P_\eta$ is not a good invariant. 
 Alternatively, in cases where a choice of helicities to construct $P_\eta$ may not be apparent, the weighted Birkhoff average could be applied to radial position, or another approximately smooth observable.
 % the method therefore requires a smooth observable, though not an invariant one. 
 In this study, $P_\eta$ proved sufficient for all considered configurations.
\section{Kinetic Poincaré sections}
\label{sec:poincare}
As discussed in previous sections, most Poincaré sections take advantage of possible dimensionality reductions for different classes of particles. In this section, we discuss the assumptions of the Poincaré map for trapped particles, as well as passing particles in a background field perturbed by a SAW.

\subsection{Poincaré sections for trapped particles}
A Poincaré map is constructed for particles trapped in one well of the equilibrium by plotting the bounce points of the particle, the points at which the parallel velocity vanishes \citep{chambliss_fast_2025, Albert_alpha_transport}. QS is not necessary to make this map since no geometrical assumptions are made. 

%Throughout, we assume the total energy of the particle is the alpha birth energy, 3.5 MeV.  

This map assumes all particles to be trapped in a single well. Consequently, banana trapped particles (trapped in the primary well) and ripple trapped particles (trapped in a subdominant well) cannot be visualised together \citep{paul_energetic_2022}, and particles that transition between classes are excluded. Similarly, particles that are barely trapped (transiting multiple field periods but bouncing at least once) cannot be captured by this method. The operational criteria for each orbit class are given in Section~\ref{sec:orbit_class}.

\subsection{Poincaré sections for passing particles with a SAW}

Kinetic Poincaré plots can be constructed for a single-harmonic SAW at fixed effective energy, provided the equilibrium is perfectly quasisymmetric.
% Each particle is initialized at a constant $E^\prime$, and energy is no longer fixed. 
We use these plots primarily to calibrate the WBA with time-independent and time-dependent magnetic perturbations.

These maps can only be constructed for a single harmonic, defined by $ \delta \Phi(\psi) \sin (\omega t+m \theta-n \zeta)$ for a single set of $m$, $n$, and $\omega$. Under this single-harmonic assumption, we define parameters $m^\prime$ and $n^\prime$, 
\begin{equation}
m^{\prime}=\frac{m N^{\prime}-n M^{\prime}}{M N^{\prime}-M^{\prime} N}, \quad n^{\prime}=\frac{m N-n M}{M N^{\prime}-M^{\prime} N}, 
\end{equation}
in terms of the toroidal and poloidal quasisymmetric helicities. 

The construction of these maps depends on the canonical momentum $P_\eta$ and the energy
\begin{equation}
    E(\psi, \chi, \eta, v_{\|},t) = \frac{M_i v_{\|}^2}{2}+\mu B_0(\psi, \chi)+q \delta \Phi(\psi) \sin \left(\omega t+m^{\prime} \chi-n^{\prime} \eta\right).
\label{eq:E}
\end{equation}
Along a particle trajectory, the combination
\begin{equation}
    n^{\prime} E-\omega P_\eta=E^{\prime}
    \label{eq:Eprime}
\end{equation}
is conserved. 
%The parallel velocity for each particle is calculated at a fixed $E^\prime$, by root solving Equation \ref{eq:Eprime} substituting in Equation \ref{eq:E} and Equation \ref{eq:peta}.
We use this conserved quantity to calculate the parallel velocity on a constant $E'$ surface, assuming constant $\mu$, specifying $\operatorname{sign}(v_{\|})$, and specifying initial spatial conditions. FIRM3D advances the guiding-centre orbits until they intersect a plane of constant $\eta-\omega t/n'$, thus suppressing the remaining explicit phase dependence. With QS error, the effective energy is no longer invariant, and the Poincaré maps are therefore higher-dimensional. To circumvent this, quasisymmetry error can be suppressed in FIRM3D’s magnetic field interpolant, enabling the construction of these kinetic Poincaré maps.

\subsection{Computing SAW resonant locations}

As derived by \cite{alex}, a general resonance condition for passing particles in quasisymmetric (QS) configurations is given by
\begin{equation}
    \Omega= \omega + (m' + \ell) \omega_\chi - n' \omega_\eta = 0,
    \label{eq:resonance_condition}
\end{equation} where $\omega$ is the mode frequency, $\ell$ is an integer arising from helical drift couplings, and  $\omega_\chi\text { and } \omega_\eta$ are the transit drift frequencies. We numerically compute the orbit-averaged frequencies $\omega_\chi (\mu, P_\eta, E)$ and $\omega_\eta (\mu, P_\eta, E)$ in the equilibrium magnetic field. 
This allows us to determine the resonant locations in $P_\eta$ for each value of $\ell$ satisfying \eqref{eq:resonance_condition}. %Alternatively, the frequencies can be evaluated as functions of $(P_\eta, \mu, E)$. 

For a single harmonic, $n^\prime$ is fixed so that the resonance condition has $m^\prime + \ell$ dependence, leading to well-separated resonances. In tokamaks, AE harmonics typically share toroidal mode values $n$, so the resonance condition solely has $m + \ell$ dependence \citep{Mynicklown}. As a result, all resonances of a given AE reside on a constant $E^\prime$ surface, and transport is confined to this surface. 

In contrast, the MHD vorticity equation couples $n$ values in a stellarator through the number of field periods,
% Magnetic field period coupling is introduced to the mode numbers, 
resulting in eigenmodes with harmonics of many $(n, m)$ values. 
% Two harmonics in an eigenmode with distinct $(n, m)$ values can share a $n^\prime$, and therefore share a value of $E^\prime$. 
Therefore, for a given AE, the resonance conditions for each harmonic are dependent on $n^\prime$ and  $m^\prime + \ell$. As $n^\prime$ is not fixed for all harmonics, the harmonics of an AE do not necessarily share a common $E^\prime$ surface, as they do in tokamaks, and no single two-dimensional section captures the resonant transport. %Additionally, the MHD equations couple $n^\prime$ values in a stellarator through the structure of the magnetic field itself. Considering the flute-like assumption ($k_{\|} / k_{\perp} \ll 1$) of the Alfvén vorticity equation 
%$$\omega^2 \nabla \cdot\left(\frac{1}{v_{\mathrm{A}}^2} \nabla \delta \Phi\right)+B_0 \nabla_{\|}\left[\frac{1}{B_0} \nabla_\perp^2\left(\nabla_{\|} \delta \Phi\right)\right]=0$$
%with Alfvén speed $v_A = B_0 (s, \theta, \zeta) / \sqrt{\mu_0 \rho_0}$, and mass density $\rho_0$, that takes solutions of the form $\delta \Phi=\sum_{m, n} \Phi_{m, n}(s) e^{i\left(m \theta-n \zeta+\omega t+\varphi_{m, n}\right)}$ (\cite{spongstability}), a field period coupling is introduced to the mode numbers through the magnetic field terms, and the Jacobian when evaluating the derivatives.  This indicates that $n^\prime$ is coupled in stellarators in a way that is not true for axisymmetric tokamaks. 

%It is also important to note that no coordinate system can be constructed such that particle trajectories can be transformed to 2D space for non-QS stellarators, such as QI configurations. Because of this, Poincaré maps cannot be made for QI equilibria with an AE.

\section{Weighted Birkhoff averaging}
\label{sec:WBA}
% Integrable motion is periodic or quasiperiodic.  Chaotic motion is irregular and has large differences in final states despite having similar initial states. 
In an integrable Hamiltonian system, bounded orbits are confined to invariant tori. Upon sufficiently small perturbation, such as that of a small-amplitude SAW,
% or one whose frequency lies close to a resonance with the unperturbed orbital frequencies, 
KAM theory shows that tori with sufficiently irrational frequencies persist, while others are destroyed and replaced by isolated periodic orbits, island chains, and chaotic regions \citep{kolmogorov1954,arnold1963proof,moser1962}. 

The weighted Birkhoff average (WBA) quantifies chaos in flows and converges more rapidly than standard estimators of the maximal Lyapunov exponent \citep{meissdun,sander2020birkhoff}. It is defined as the time average of a function evaluated along a flow integrated against a bump function \citep{meissdun}, 
\begin{equation}
\text{WBA}_T(p)= \int_0^T C e^{-\frac{1}{ \left(\frac{t}{T}\left(1-\frac{t}{T}\right) \right)}} p(t) d t,
\label{eq:WBA}
\end{equation}
where $p(t)$ is the observable evaluated along the trajectory, $t/T$ is the normalised flow time, $T$ is the final time, and $C$ is a normalising factor $$C = \left( \int_0^T e^{-\frac{1}{\left(\frac{t}{T}\left(1-\frac{t}{T}\right) \right)}} dt \right)^{-1}.$$  

WBA provides a powerful tool for quantifying chaos due to its super-convergent behaviour. Specifically, for quasiperiodic trajectories with Diophantine rotation vectors, or rotation frequencies far from low-order rationals, WBA converges to the time average of an observable faster than $\mathcal{O}(T^{-h})$ for any positive integer $h$ \citep{meissdun}. As a result, the relative time to WBA convergence distinguishes between regular and chaotic trajectories. WBA convergence is a comprehensive alternative to Poincaré maps, since it does not rely on any knowledge of particle behaviour (such as trapping class) or any underlying symmetries. This means it can indicate when perturbations become chaotic even when sufficient constants of motion do not exist to make the necessary dimensionality reductions for Poincaré maps.

We quantify convergence by the number of converged digits, comparing the average at time $T$ with that at $T/2$, inspired by the scheme of \cite{meissdun}. This is normalised in the expression for the relative digit accuracy, 
\begin{equation}
\operatorname{relative\ digit\ accuracy}_T(p) \equiv-\log _{10} \left( \frac{\left|\mathrm{WBA}(p)_{T / 2}-\mathrm{WBA}(p)_T\right|}{\frac{1}{2}\left(\left|\mathrm{WBA}(p)_{T / 2}\right|+\left|\mathrm{WBA}(p)_T\right|\right)} \right).
\label{eq:relDA}
\end{equation}

The relative digit accuracy fails when the average value for $p$ approaches zero, as the denominator becomes small. To compensate for this, we select the final digit accuracy as the maximum of the $\operatorname{relative\ digit\ accuracy}_T(p)$ and the $\operatorname{local\ digit\ accuracy}_T(p)$, expressed by 

\begin{equation}
\operatorname{local\ digit\ accuracy}_T(p) \equiv-\log _{10} \left( \frac{\left|\mathrm{WBA}(p)_{T/2}-\mathrm{WBA}(p)_T\right|}{\max_{t \in (0,T]} \left(\left|p\left(t\right)\right|\right)} \right). 
\end{equation}
This normalises the absolute digit accuracy, $ -\log _{10} \left|\mathrm{WBA}(p)_{T/2}-\mathrm{WBA}(p)_T\right|$, to remain well defined near zero and remain independent of the scale and units of $p(t)$. This approach is analogous to the methodology of \cite{meissdun}, in which the digit accuracy is taken as the maximum of the absolute, $-\left( \log _{10}\left|\mathrm{WBA}(p)_{T/2}-\mathrm{WBA}(p)_T\right| \right)$, and the relative digit accuracy, as defined in \eqref{eq:relDA}. We depart from their convention to ensure the digit accuracy is always measured against the largest magnitude $p(t)$ attains rather than against a vanishing mean. The digit accuracy utilised is 
\begin{equation}
\operatorname{digit\ accuracy}_T(p) \equiv \max \left ( \operatorname{local\ digit\ accuracy}_T(p), \operatorname{relative\ digit\ accuracy}_T(p) \right ).
\label{eq:DA}
\end{equation}

 We integrate guiding-centre trajectories with FIRM3D, using the Landreman-Buller $\beta=2.5\%$ QA (LBQA) equilibrium \citep{landreman2022optimization}, with a single-harmonic SAW calculated in AE3D \citep{spongstability}. FIRM3D can enforce exact quasisymmetry by suppressing the symmetry-breaking field harmonics. This permits perfect conservation of the effective energy and construction of a Poincaré plot, which would not be the case for an imperfectly quasisymmetric stellarator.  To assess the digit accuracy and to fix the two free parameters, the integration time $T$ and the chaos threshold, we evaluate the relationship between visible structures in a Poincaré map and the digit accuracy calculated through WBA. 

\begin{figure}
    \centering
        \includegraphics[width=\textwidth]{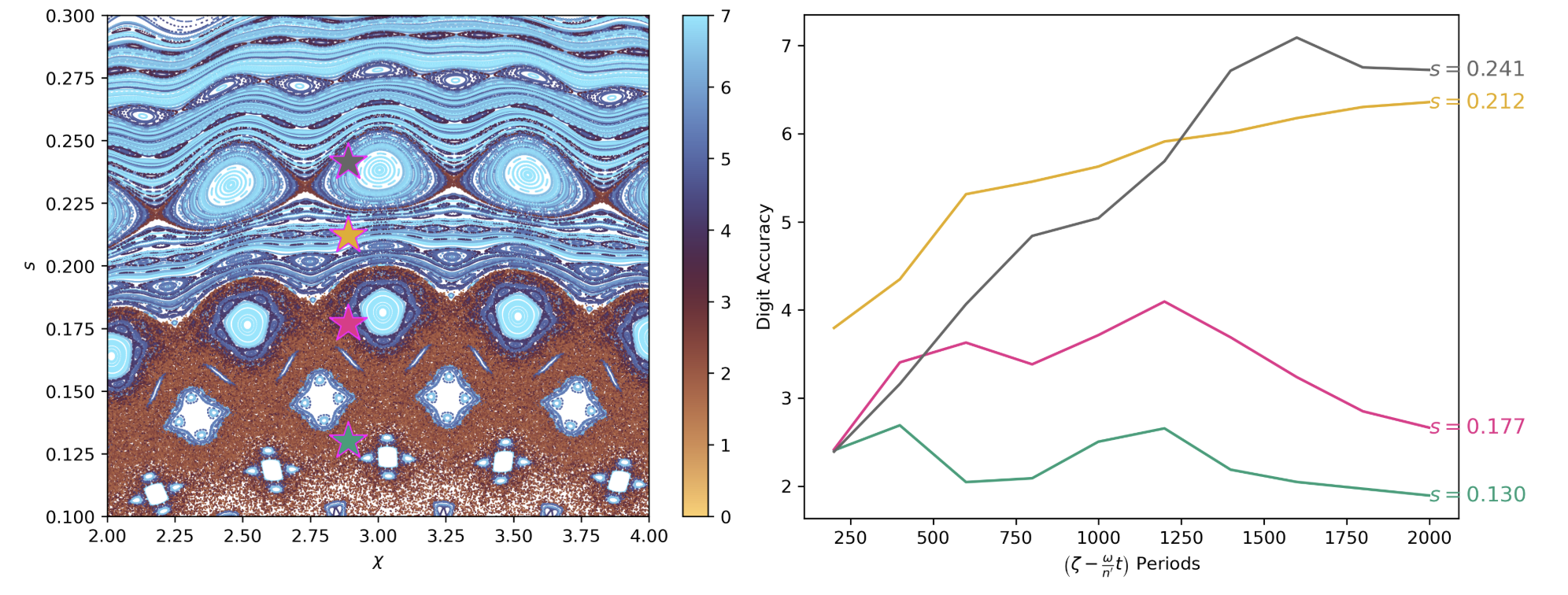}
        \caption{Left: Poincaré plot with a single $(m, n) = (11, 4)$ harmonic of uniform radial amplitude in the LBQA equilibrium with enforced QS, coloured by digit accuracy. Right: digit accuracy as a function of integration time, measured in periods of $\zeta - \omega t/n^\prime$, for the four initial conditions marked by stars in the left panel; curve colours match the star colours.} 
        \label{fig:poinc_chaos}
\end{figure}

A Poincaré map with a single $(m, n) = (11, 4)$ harmonic in a perfectly enforced QA background field is shown in Figure \ref{fig:poinc_chaos}. The digit accuracy is evaluated at $T = 1 \times 10^{-2} $ s for all maps. The integration time was chosen by plotting the digit accuracy as a function of $T$, as in Figure \ref{fig:poinc_chaos}, and evaluating the time to convergence for integrable and non-integrable particles that can be distinguished in the map. All passing particles reach a digit accuracy plateau within 1250 toroidal transits, which is of order $10^{-2}$ s for passing particles in the LBQA. Similar convergence rates are observed more generally, including for the LBQA trapped particle Poincaré map without enforced quasisymmetry shown in Figure \ref{fig:trapped_chaos}, as the non-chaotic layers of trapped particles converge by $10^{-2} $ s.

There are four regions of interest in Figure \ref{fig:poinc_chaos}: inside an intact island (indicated in grey), the stochastic region of island overlap (indicated in green), the boundary layer where particles fall in and out of chaotic layers (indicated in magenta), and the intact drift surfaces around the regions of overlap (indicated in yellow). The digit accuracy reveals when a particle drifts into and out of the resonant region. An example of this is indicated in magenta, which sits on the boundary between the stochastic region and the intact island in the Poincaré map, with a digit accuracy which increases and then falls. A clear boundary of digit accuracy can be seen between chaotic regions and regions with intact drift islands and surfaces, as seen in the gap between the grey/yellow and magenta/green pairings. This contrast allows the digit accuracy to serve as a practical numerical diagnostic for differentiating chaotic and integrable motion. 
% There exists a clear alignment between regions of low digit accuracy and regions of broken drift surfaces visible in the Poincaré map. 
However, in order to use digit accuracy as a quantitative classifier for particle motion, it is necessary to define a threshold value that separates these dynamical regimes.

\begin{figure}
    \centering
    \begin{subfigure}{0.49\textwidth}
        \centering
        \includegraphics[width=\textwidth]{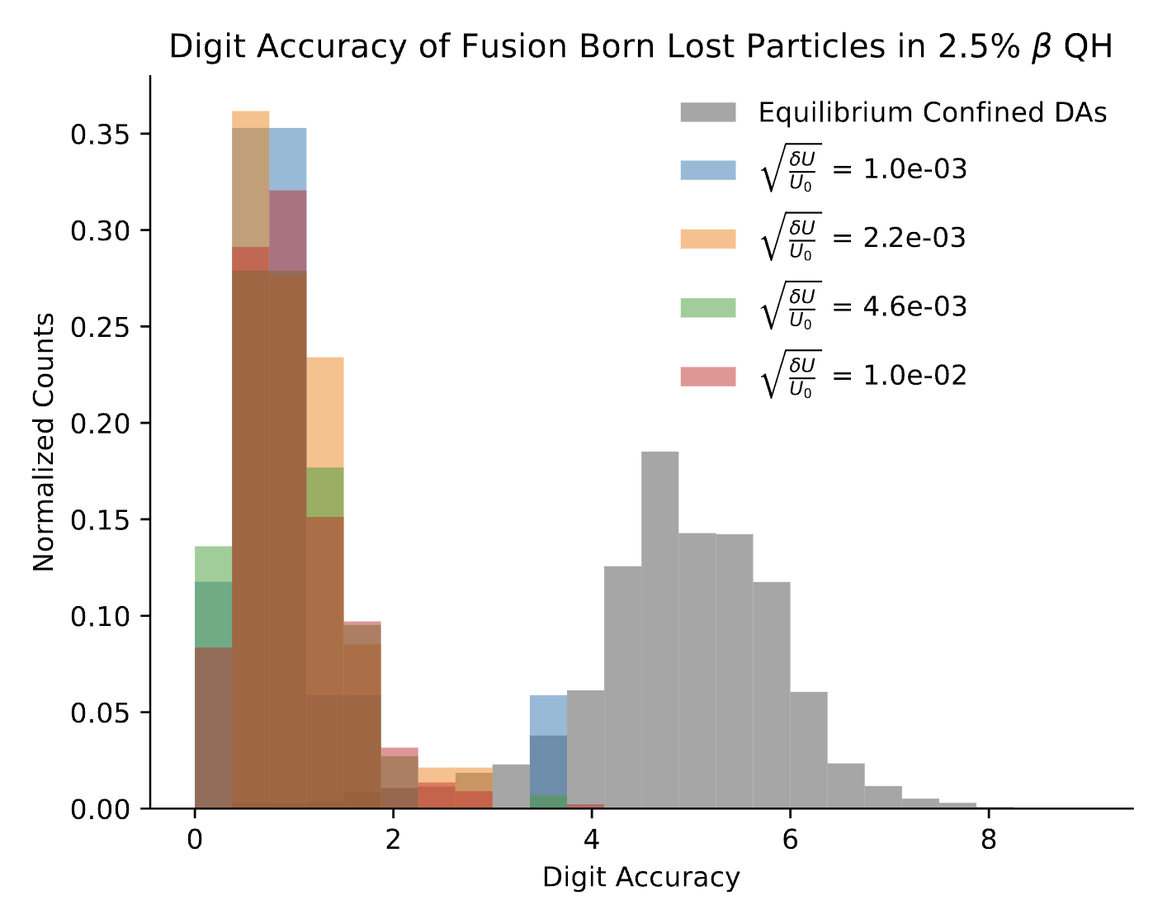}
        \caption{A 10-harmonic AE in the LBQH \citep{landreman2022optimization}.}
    \end{subfigure}
    \hfill
    \begin{subfigure}{0.49\textwidth}
        \centering
        \includegraphics[width=\textwidth]{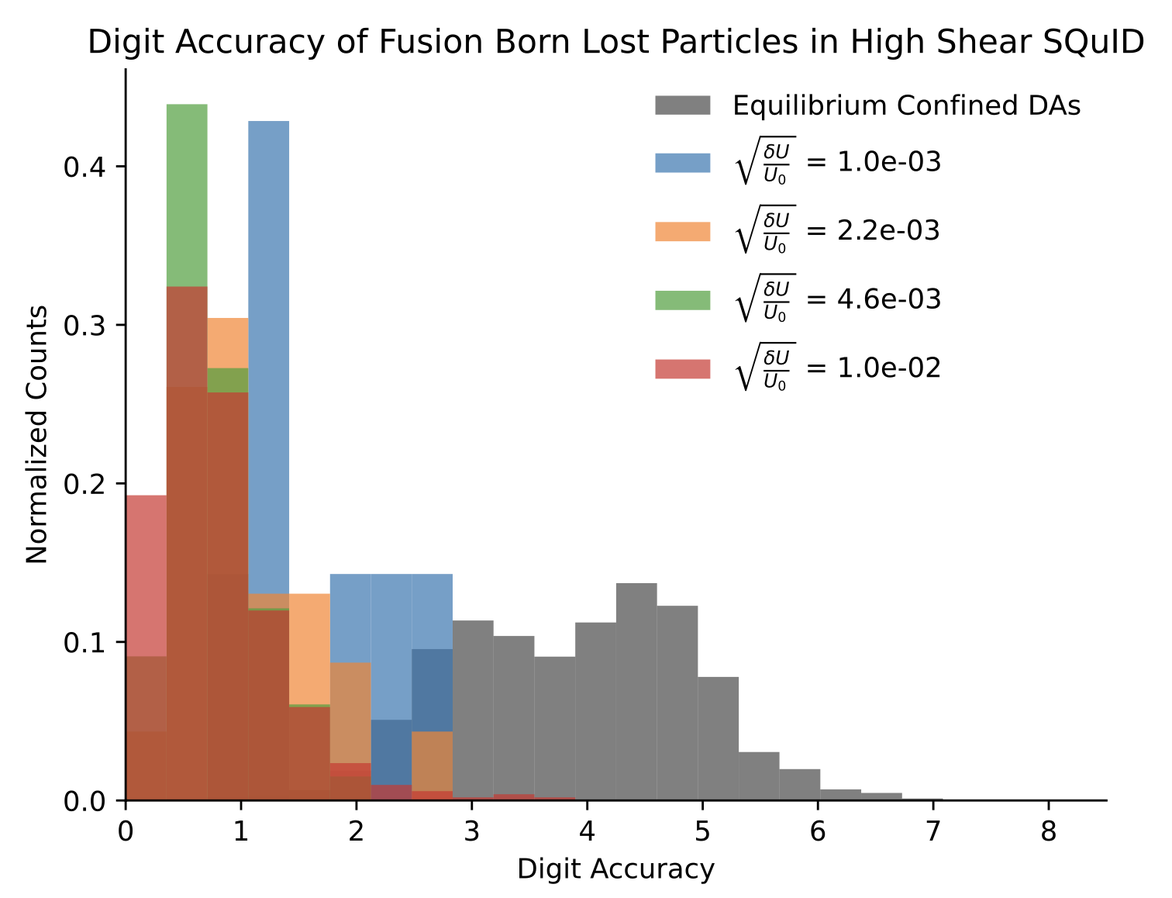}
        \caption{A 20-harmonic AE in high-shear SQuID \citep{goodsquid,alex}.}
    \end{subfigure}
    \caption{Histograms of the digit accuracy for lost particles. The digit accuracies of lost particles lie almost entirely below 3, which informed the threshold for the digit accuracy. }
    \label{fig:sidebyside}
\end{figure}

To motivate the choice of threshold digit accuracy to distinguish between chaotic and integrable behaviour, we apply AE harmonics of increasing strengths in several equilibria. Particles are spatially initialised according to a fusion birth profile proposed by \cite{bader2021modeling}. The parallel velocity is chosen uniformly between values of $v_\| = -\sqrt{2 E/M_i}$ and $v_\| = \sqrt{2 E/M_i}$. 
% We remove any particles lost (hits $s=1$) in the equilibrium with zero AE perturbation, as it is possible to lose particles through drift surfaces that intersect the reactor wall or through chaos in the equilibrium field transporting particles towards $s=1$. 
The AE magnitude is quantified using the perturbed energy, 
\begin{align}
\frac{\delta U}{U_0}=\frac{\int d\mathbf{x} \,\delta B^2}{\int d\mathbf{x} \, B_0^2}.
\label{eq:perturbed_energy}
\end{align}
We then isolate particles lost as a result of the AE by excluding the equilibrium losses. By removing equilibrium losses, we isolate particle trajectories whose loss must occur due to chaotic dynamics, since these particles are no longer confined to an intact torus and are therefore non-integrable. This selection is therefore conservative, as it may miss chaotic particles lost in the equilibrium, but it will not admit regular ones. The particles lost in the presence of large AEs are lost as a result of KAM surfaces breaking, so these particles are deemed chaotic.
% , even if some chaotic particles may remain well confined by a transport barrier. 

Figure \ref{fig:sidebyside} shows representative digit accuracy distributions for lost particles in two distinct equilibria perturbed by an AE: the Landreman-Buller $\beta=2.5\%$ QH (LBQH) and high-shear SQuID configuration introduced in \cite{landreman2022optimization} and \cite{goodsquid,alex}, respectively. In both cases, the overwhelming majority of lost particles exhibit digit accuracy values below 3. As with any binary threshold, there is a possibility of false positives and false negatives. Notably, this trend is robust across both equilibria, suggesting that the stochasticity associated with particle loss is not tied to a specific configuration but instead reflects a more general property. The distribution of the lost particles motivated our choice of a digit accuracy of 3 as the operational threshold between chaotic and integrable dynamics for all equilibria going forward. Although the precise numerical value may vary slightly with the equilibrium geometry, the qualitative transition is robust across the studied configurations. 

\begin{figure}
    \centering
    \includegraphics[width=0.6\linewidth]{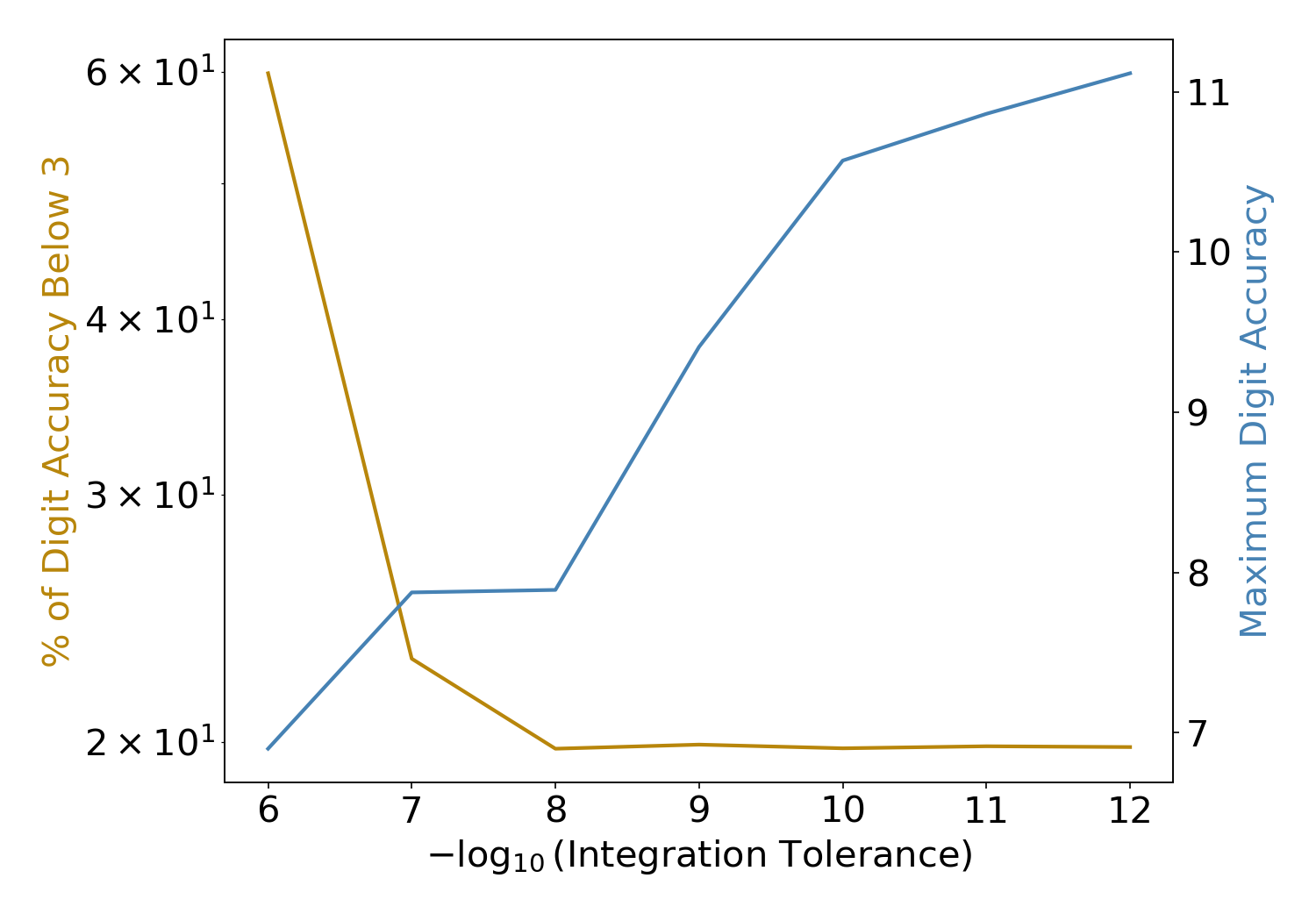}
    \caption{Maximum digit accuracy and percentage of particles with digit accuracy below the threshold of 3 as a function of integration tolerance.}
    \label{fig:integration}
\end{figure}

 The maximum achievable digit accuracy is fundamentally bounded by the tolerance of the ODE solver used to integrate the equations of motion defining the flow.
 As seen in Figure \ref{fig:integration}, although the maximum digit accuracy increases as the integration tolerance is tightened, the proportion of particles with a digit accuracy below the threshold of 3 saturates beyond a tolerance of $10^{-8}$. This suggests that the classification of chaotic trajectories by this metric becomes insensitive to further reductions in integration tolerance. In this study, we maintain a tolerance of $10^{-10}$. 
 % Despite digit accuracy dependence on integrator tolerance, we find that the threshold remains resilient for a range of possible integration tolerances, as seen in Figure \ref{fig:integration}. 

 \cite{meissdun} chose a threshold of about 5. However, their threshold was sensitive to the equations of motion. For example, a dynamical system with a chaotic attractor required a non-binary classification scheme. They chose their chaotic threshold as the observed separation in digit accuracies between the visibly chaotic and non-chaotic trajectories when comparing the digit accuracy as a function of integration time. The bulk of trajectories reported as chaotic in that work have digit accuracies similar to chaotic particles in this work, which remain almost entirely below 3. 
 However, the maximum digit accuracy reported by \cite{meissdun} was $\approx 15$, using absolute and relative integrator tolerances between $10^{-10}$ and $10^{-15}$, depending on the equations of motion.  The maximum digit accuracy of regular trajectories depends on the integration tolerance (as seen in Figure \ref{fig:integration}), the choice of $T$, and the system being integrated. The larger maximum digit accuracy reported in that work would widen the gap between chaotic particle digit accuracy and regular populations whose accuracy varies based upon equations of motion and integrator tolerance. In this work, the maximum digit accuracy attained by regular orbits is lower (Figure \ref{fig:integration}) than in \cite{meissdun}. Since chaotic orbits cluster near 2, a threshold of 3 better separates the two populations than the value of 5.  Additionally, we are primarily interested in strongly chaotic trajectories which would drive transport, motivating a deliberately conservative choice of threshold digit accuracy. Therefore, their larger classification threshold does not contradict this study. 

\subsection{Phase-space maps}
 \label{phase_maps}

This work introduces WBA-based phase-space maps. We construct the map with test particles initialised throughout the accessible phase space given a reference energy and magnetic moment. The digit accuracy is averaged over particles binned by invariants that would define a KAM surface in the unperturbed state. This averaged metric quantifies chaotic radial transport for possible locations on the surface. For the considered omnigeneous stellarators, we construct these maps for marker particles initialised on a grid of $\mu$ and $s$. For each $\mu$, $s$ combination, particles are uniformly initialised in $\chi$ and $\eta$. 

Maps in this work are instantiated on a grid resolution of 35 $\mu$ values, 35 $s$ values, and 25 particles for each $(s, \mu)$ pair, unless otherwise stated. The particles are traced for $ 1 \times 10^{-2} $ s, and the digit accuracies for particles are binned according to their initial $\lambda = \mathrm{sign}(v_\parallel) \mu B_{\text{min}}/E$ and $P_\eta$ values, where $B_{\text{min}}$ is the minimum field strength in the volume and $E$ is the total energy.  The particle's digit accuracy is averaged within $\left( \lambda, P_\eta \right)$ bins. We average over bins defined by initial $\lambda$ and $P_\eta$, even in cases where these variables are not perfectly conserved, to demonstrate which initial conditions lead to chaotic behaviour. For QI stellarators, we average over bins in $\left( \lambda, s \right)$.
% $B_{\text{min}}$ is the minimum field strength in the volume, and $E$ is the reference energy, $E =$ 3.5 MeV. All phase space plots are binned in $\lambda = \text{sign}(v_{\|}) \frac{\mu}{E} B_{\text{min}}$ and initial $P_\eta$ value.  

In the equilibrium (unperturbed) case, the initial parallel velocities are calculated at a fixed reference energy $E =$ 3.5 MeV. In the case with a single-harmonic SAW, the particles are initialised with parallel velocity calculated at a fixed $E^\prime$ determined by the $n^\prime$, $m^\prime$ and $\omega$ values of the harmonic, using \eqref{eq:Eprime}. For maps with a multi-harmonic AE, the initial parallel velocity is calculated at the $E^\prime$ of the largest-energy harmonic as defined by \eqref{eq:perturbed_energy}. Note that the particle energy is no longer fixed, although $\lambda$ is defined with respect to the reference energy $E =$ 3.5 MeV. For phase-space maps with a SAW, particles lost in the equilibrium are removed, as most of these have drift surfaces that intersect the wall and are not traced long enough to compute a valid digit accuracy. 
\section{Departures from integrability without SAWs}
\label{sec:equilibrium_chaos}

\begin{figure}
    \centering
        \includegraphics[width=0.6\textwidth]{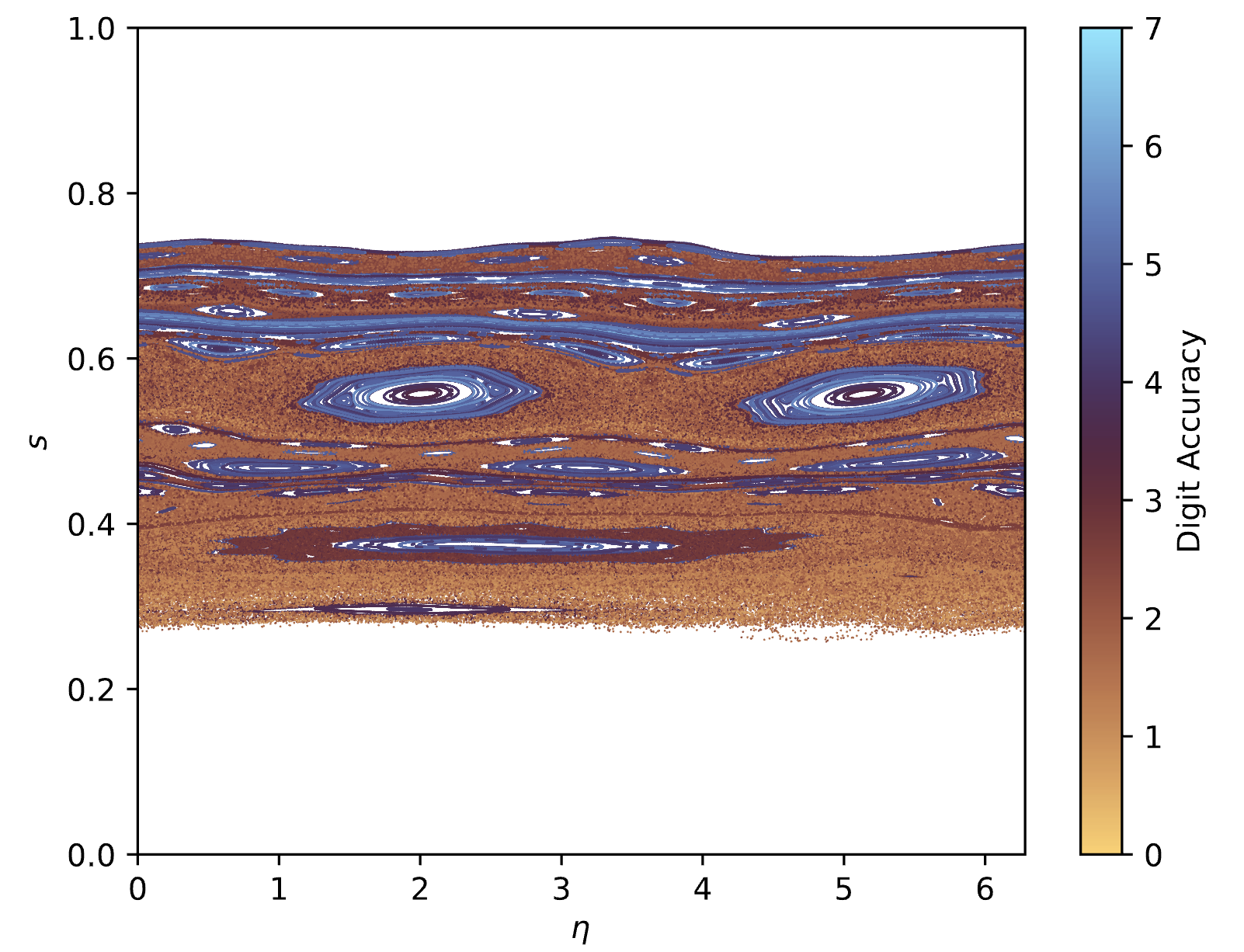}
        \caption{Poincaré plot of trapped particles 
        with 
        % with a mirror point of $(s, \theta, \zeta) = (0.5, \pi/2, 0)$ at a 
        $\lambda = 0.91$ 
        % $\lambda=\text{sign}\left(v_\| \right) \frac{\mu}{E} B_{\text{min}}$, 
        in the unperturbed LBQA without QS enforced. The corresponding location of this cross section is highlighted with the yellow line in Figure \ref{fig:QSerrormapping}. } 
        \label{fig:trapped_chaos}
\end{figure}

The LBQA has a significant population of chaotic trapped particles, which \cite{chambliss_fast_2025} attribute to QS error in the equilibrium. These particles fall into several classes, including barely trapped, ripple trapped, and banana trapped. A Poincaré map for particles trapped in the main well is shown in Figure \ref{fig:trapped_chaos}. 

% Visualizing the barely trapped and ripple trapped particles is important for understanding which regions of phase space behave stochastically, and cannot be done with the trapped map. 
Unlike the trapped map, the phase-space map enables a quantification of chaos for barely trapped and ripple trapped particles.
To construct the phase-space map, we follow the procedure detailed in Section \ref{phase_maps}. Each vertical slice in $\lambda$ of the phase-space map corresponds to the averaged digit accuracy as a function of $P_\eta$. In this way, each vertical $\lambda$ slice of the phase-space map represents the same trajectories portrayed in a Poincaré map; however, the metric is averaged over the angles. 

\begin{figure}
    \includegraphics[width=\textwidth]{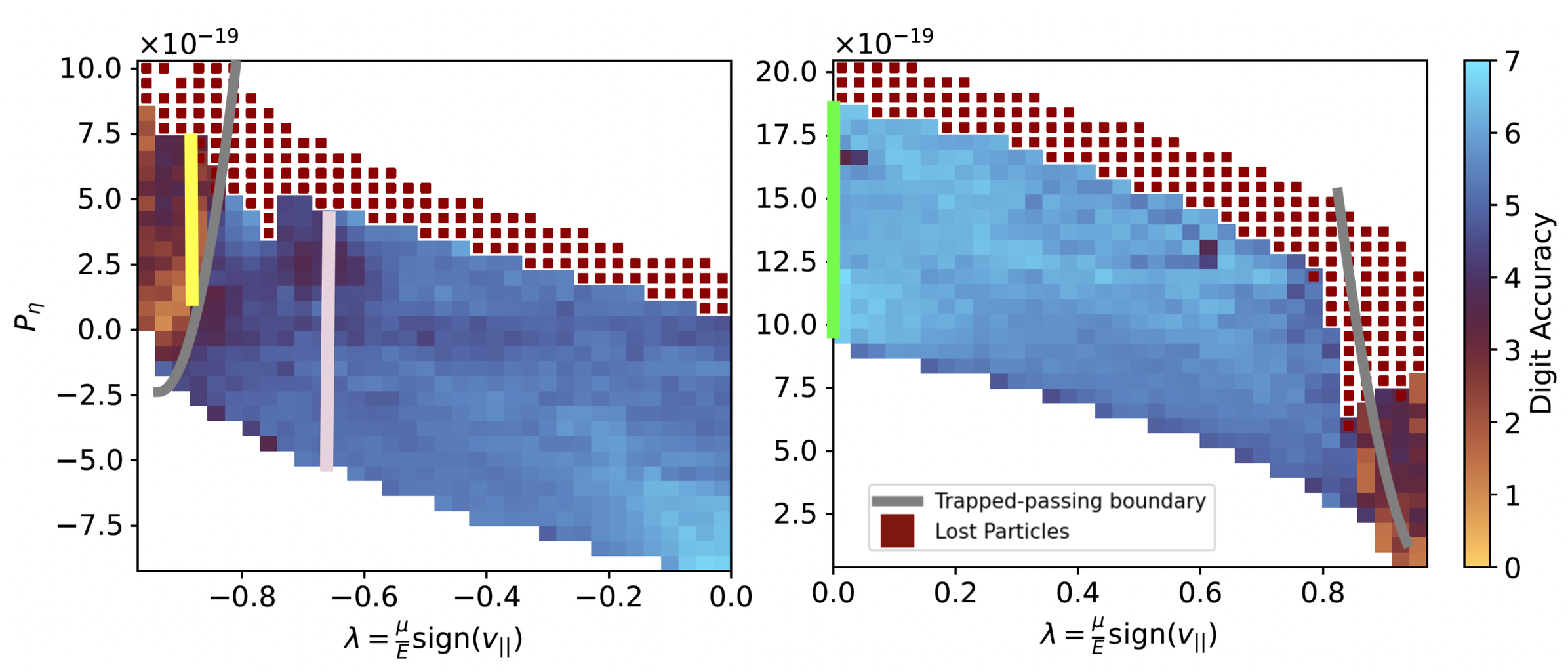}
    \caption{Digit accuracy as a function of canonical momentum $P_{\eta}$ and trapping parameter $\lambda$. The grey line is the passing/trapped boundary. Red markers are placed over regions with lost particles. The location of the yellow line corresponds to the $\lambda=0.91$ Poincaré map shown in Figure \ref{fig:trapped_chaos}, the pink line to the $\lambda=-0.75$ Poincaré map in Figure \ref{fig:QA_lam-0.75}, and the green line to the $\lambda=0.0$ Poincaré map in Figure \ref{fig:QA_lam0.0}. A total of 30,625 particles (35 sampled $\mu$ points, 35 sampled $s$ points, and 25 uniformly distributed points in $\chi$, $\eta$ on each $s$) were traced in the unperturbed LBQA without QS enforced.}
    \label{fig:QSerrormapping}
\end{figure}

\begin{figure}
    \centering

        \begin{subfigure}{0.482\textwidth}
        \centering
            \includegraphics[width=\linewidth]{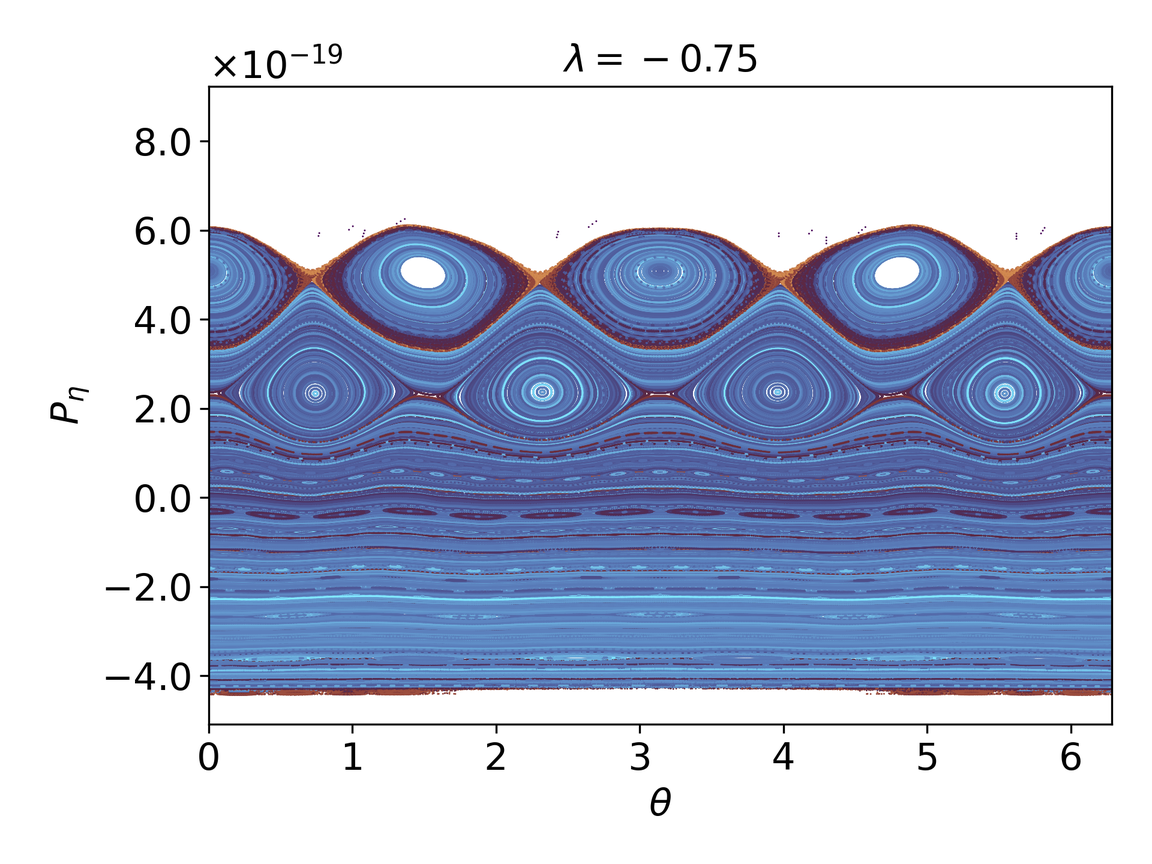}
            \caption{Corresponds to the pink line in Figure \ref{fig:QSerrormapping}.}
            \label{fig:QA_lam-0.75}
        \end{subfigure}
        \begin{subfigure}{0.482\textwidth}
        \centering
            \includegraphics[width=\linewidth]{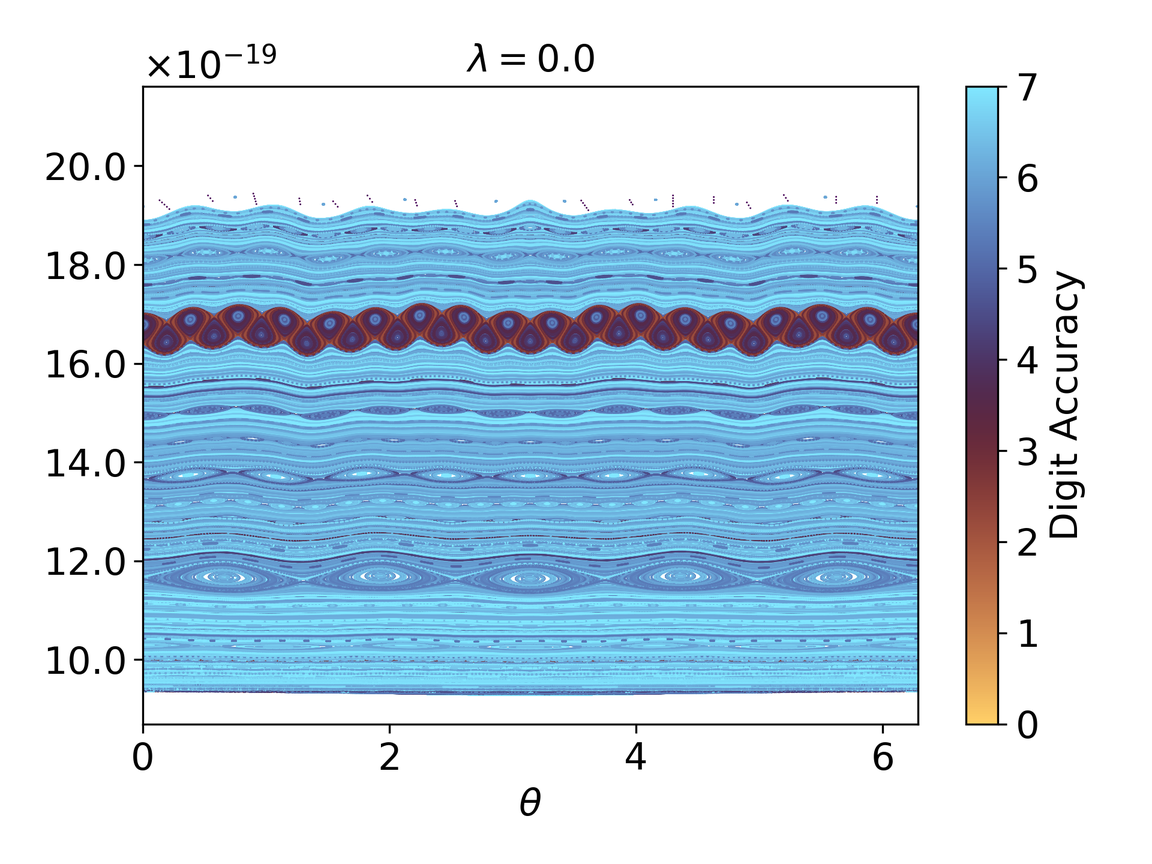}
            \caption{Corresponds to the green line in Figure \ref{fig:QSerrormapping}.}
            \label{fig:QA_lam0.0}
        \end{subfigure}
    
    \caption{Passing Poincaré maps for the LBQA equilibrium without SAWs.
    % Note the y-axis is $P_\eta$ rather than flux surface, $s$. 
    }
\end{figure}

The chaotic regions for trapped particles are evident in Figure \ref{fig:QSerrormapping}, with the transition from integrable to non-integrable nearly coinciding with the passing/trapped boundary. 
% This is similar to what is visible in the trapped particle Poincaré map, but with greater flexibility to also visualize the barely and ripple trapped particles. 
Regions of declining digit accuracy for passing particles are noticeable in Figure \ref{fig:QSerrormapping}, such as along the pink line. This region can be investigated with a Poincaré map for passing particles at $\lambda = -0.75$, as seen in Figure \ref{fig:QA_lam-0.75}. 

In Figure \ref{fig:QSerrormapping}, an island chain grows in drift space as the magnitude of $\lambda$ increases. Around $\lambda=- 0.75$, the island becomes sufficiently large to experience separatrix breaking, matching the localised decrease in digit accuracy in that region of Figure \ref{fig:QSerrormapping}.  Figure \ref{fig:QA_lam0.0} demonstrates a region of island-overlap-induced stochasticity, which is reflected in the full phase-space map. 

By evaluating the resonance condition \eqref{eq:resonance_condition} with a static ($\omega=0$) perturbation, we find that the islands in Figure \ref{fig:QA_lam-0.75} correspond to the rational $h = 1/4$ resonance. Additionally, the chaotic islands in Figure \ref{fig:QA_lam0.0} correspond to the $h=8/17$ resonance, where $h=\omega_\theta/\omega_\zeta$ is the drift helicity. This resonance is significant because it occurs near the turning point of the drift helicity for $\lambda=0$. The island chain appears twice because the drift helicity satisfies the resonance condition at two nearby locations, similar to behaviour observed in \cite{chambliss_fast_2025}.

\subsection{Chaos by orbit class}
\label{sec:orbit_class}
\begin{figure}
    \centering
            \includegraphics[width=\textwidth]{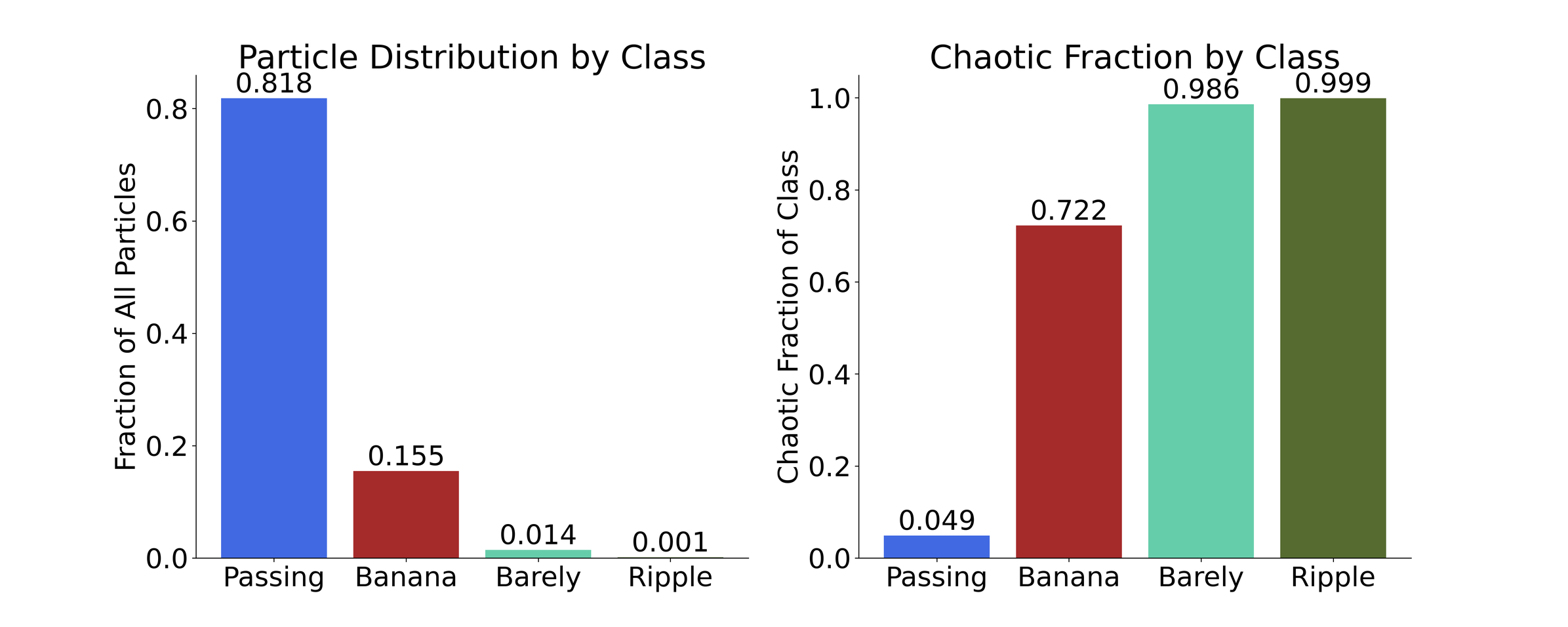}
    \caption{Distribution of confined orbit classes sampled from the fusion birth distribution \citep{bader2021modeling}, alongside the fraction of each orbit class that is chaotic (has a digit accuracy less than 3).}
\label{fig:orbitclassbreakdown}
\end{figure}

We evaluate the relationship between different classes of trapped particles in phase space and the digit accuracy using the orbit classification method of FIRM3D, as specified in \cite{FIRM3DJOSS}. We initialise 10,000 alpha particles at fixed energy according to the fusion-born distribution of \cite{bader2021modeling}, classify each orbit, and compute the corresponding WBA. Each particle is characterised by the fraction of its bounce segments falling in each trapped category. A particle with no bounce segments is classified as passing. Bounce segments are defined by the change in $\chi$ between mirror points, $v_\| = 0$, along the trajectory. Each bounce segment for each particle is delineated into banana, barely, or ripple categories by comparing the actual change in $\chi$ during the bounce segment with the change expected for a particle trapped in the primary well, defined by the critical magnetic field on the surface for mirroring given the pitch angle. If the particle passes through more than 1.25 periods of the helical angle during that bounce segment, the segment is deemed barely trapped. If the particle passes through less than half of the critical helical angle, then the segment is ripple trapped. A particle is classified as passing if it remains passing for the entire trajectory, as barely trapped if any bounce segments are deemed barely trapped, as banana trapped if more than half of its bounce segments are in the dominant magnetic well, and as ripple trapped if more than half of its bounce segments are in a non-dominant well.
The distribution of confined particles from the alpha birth distribution in each orbit classification is shown in Figure \ref{fig:orbitclassbreakdown}. Confined particles are of particular interest, as these are particles that are likely to contribute to chaotic layers that may flatten phase-space gradients, rather than being promptly lost. As seen in Figure \ref{fig:orbitclassbreakdown}, we find that most trapped particles are chaotic, with the exception of 28\% of banana trapped particles being integrable. This is consistent with the low digit accuracy region past the passing/trapped boundary demonstrated in Figure \ref{fig:QSerrormapping}. Passing particles are generally regular away from drift resonances \citep{Foster}, as observed in Figure \ref{fig:QSerrormapping}.
% which is to be expected as phase-space islands only materialize near rational surfaces. 
% \cite{Foster} notes that passing particles only break from drift surfaces at resonances, for which the islands width grows slowly with decreasing shear. Additionally, \cite{Foster} found no instances of island overlap in well optimized stellarators studied due to this result.
Previous studies \citep{BeidlerTransitioning,paul_energetic_2022} have analytically and numerically shown that transitioning particles can radially diffuse due to changes in the second adiabatic invariant from separatrix breaking, which is consistent with the increases in chaos seen in Figure \ref{fig:QSerrormapping} for barely trapped particles.
\section{Effect of QS error on chaotic structures}
\label{sec:qs_error}

\begin{figure}
    \centering
    \begin{subfigure}{\textwidth}
        \includegraphics[width=\linewidth]{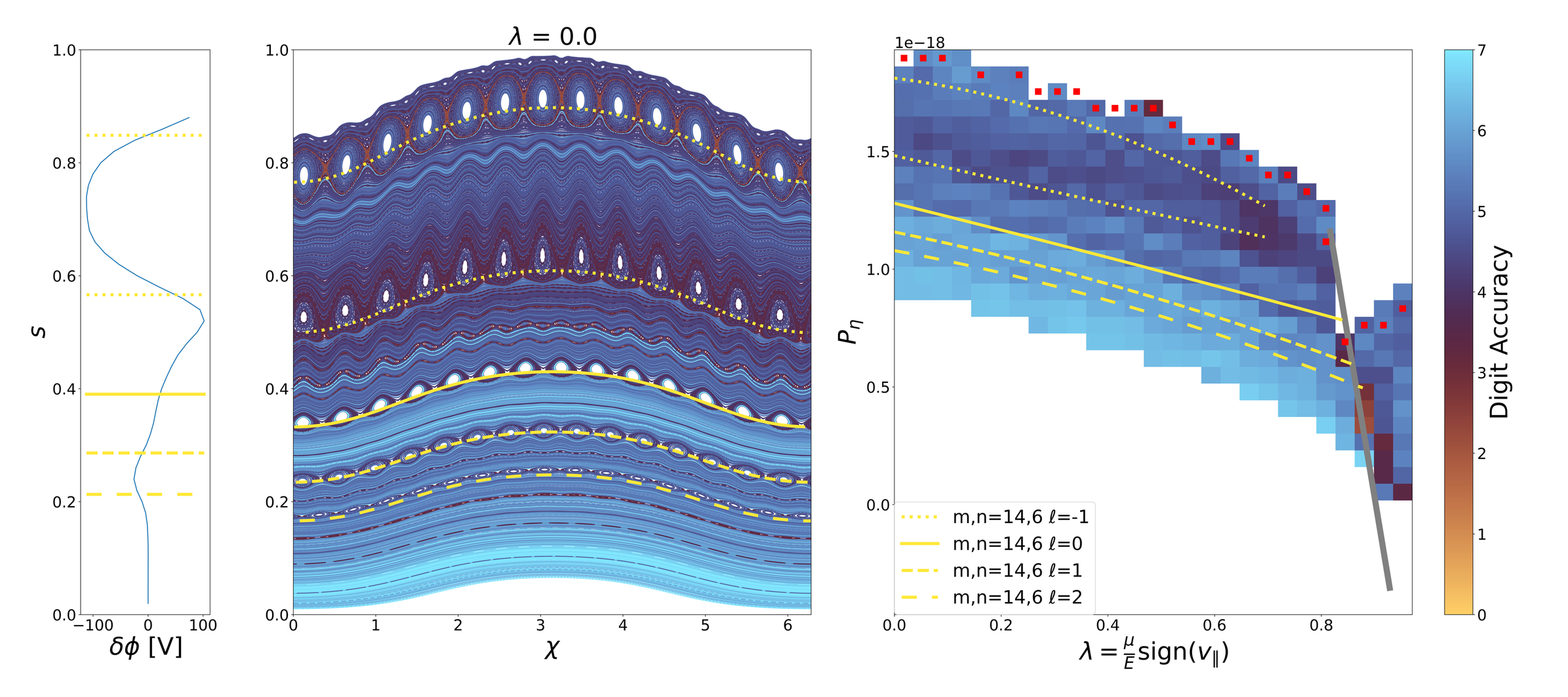}
        \caption{Full phase-space map for the LBQA with QS error suppressed.}
        \label{fig:QA4_QS}
    \end{subfigure}
    \newline
    \vspace{0.5em}
    \begin{subfigure}{\textwidth}
        \includegraphics[width=\linewidth]{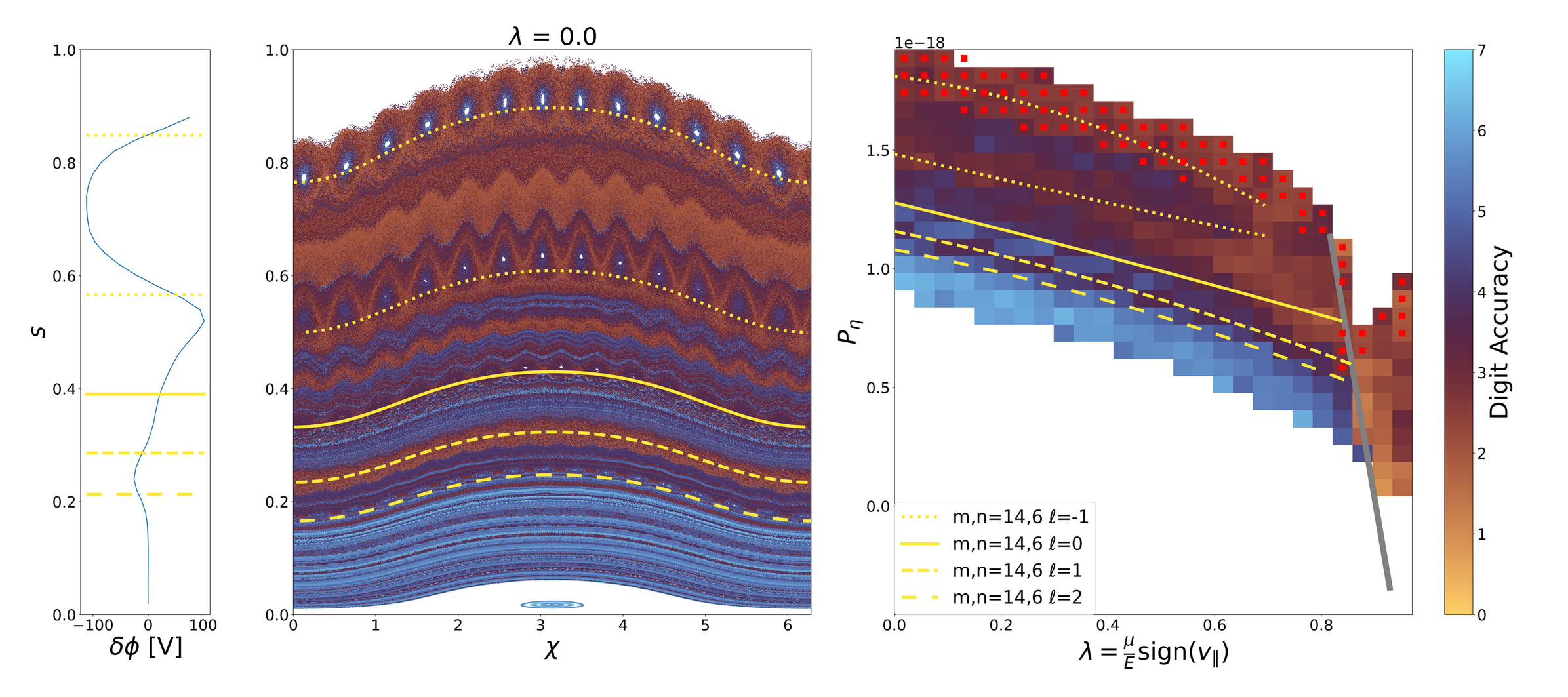}
        \caption{Full phase-space map for the LBQA without QS error suppressed. This Poincaré map is 3D due to the symmetry deviations, although it is visualised on the 2D plane.}
         \label{fig:QA4_full}
    \end{subfigure}
    \caption{The full phase-space map as compared with the Poincaré map for $\lambda=0$, co-propagating particles in the LBQA equilibrium with the $(m, n) = (14, 6)$ AE harmonic. The grey line indicates the passing/trapped boundary. Particles lost in the equilibrium field are removed, explaining the region near the passing/trapped boundary without marked digit accuracy. Regions in phase space that have at least one lost particle are denoted by a red square. Each map consists of 30,625 traced particles, specified according to Section \ref{phase_maps}. }
    %(35 sampled $\mu$ points, 35 sampled $s$ points, and 25 uniformly distributed points in $\chi$, $\eta$ on each $s$).}
    \label{fig:QApsmaps}
\end{figure}

In this section, we compare the $\lambda = 0$ Poincaré map for co-propagating particles in the LBQA with and without the suppression of QS error. We also apply the phase-space map diagnostic, sampling the digit accuracy across the full phase space in each case. We choose the largest-energy harmonic $(m, n) = (14, 6)$, which satisfies the resonance condition \eqref{eq:resonance_condition} in this equilibrium for $\ell = \pm 1$. We use a 35-harmonic AE computed for the LBQA with AE3D \citep{spongstability}, as described by \cite{alex}. The number of harmonics was truncated once the fraction of alpha particles lost to the wall converged with increasing number of harmonics, ranked by energy \eqref{eq:perturbed_energy}. The radial structure of the chosen harmonic is shown in the leftmost subplots in Figure \ref{fig:QApsmaps}.

 We enforce the appropriate perfect quasisymmetry with FIRM3D, and choose the dominant harmonic so that the $E^\prime$ invariant is well defined. In comparing the QS-enforced case to the field without suppressed error, we isolate the effects of QS imperfection on alpha-particle chaos near the resonance. 
% Widening of the chaotic layer can be noted between the perfect and imperfect QA cases. 
% The procedure for phase space map instantiation is detailed in Section \ref{phase_maps}. 

% The Poincaré map may be understood as a lower dimensional slice of the full phase space map. The Poincaré section is constructed at the fixed pitch angle, $\lambda = 0$, whereas the full phase space map calculates the digit accuracy throughout the $(P_\eta,\lambda)$ phase space volume after averaging the digit accuracy over the drift surfaces, $P_\eta$, in $\chi$ and $\eta$. The Poincaré plot corresponds to a vertical slice in the phase space map at $\lambda = 0$, retaining the orbit structure that is subsequently averaged over in the full phase space map. 

Compared to Figure \ref{fig:QA4_QS}, Figure \ref{fig:QA4_full} has a larger chaotic region of phase space, indicating that it is not only the breakdown of $P_\eta$ that contributes to regions of chaos in Figure \ref{fig:QA4_full}, but also interactions between imperfections in omnigeneity and the SAW. 
The region of stochasticity is also strongest where the harmonic has the strongest amplitude, and it coincides with the $\ell = -1$ resonance. 
The chaotic layer along the boundary of the resonant island with a broken separatrix may induce trapping and detrapping of particles, as in Figure \ref{fig:QA4_QS}.
In the perfectly quasisymmetric case with the single harmonic AE, only a fraction 0.003 of fusion-born particles are lost, as compared with the LBQA without enforced quasisymmetry, where 0.01 of fusion-born particles are lost. In the LBQA, lost particles are primarily along the resonance close to the wall. The separatrices of the resonant islands break up into a chaotic layer in the imperfectly quasisymmetric LBQA, likely through overlap with the islands generated by the QS error, and the extent of phase space in $P_\eta$ with lost particles spans the chaotic region, indicating increased transport.

\begin{figure}
\centering
\includegraphics[width=\textwidth]{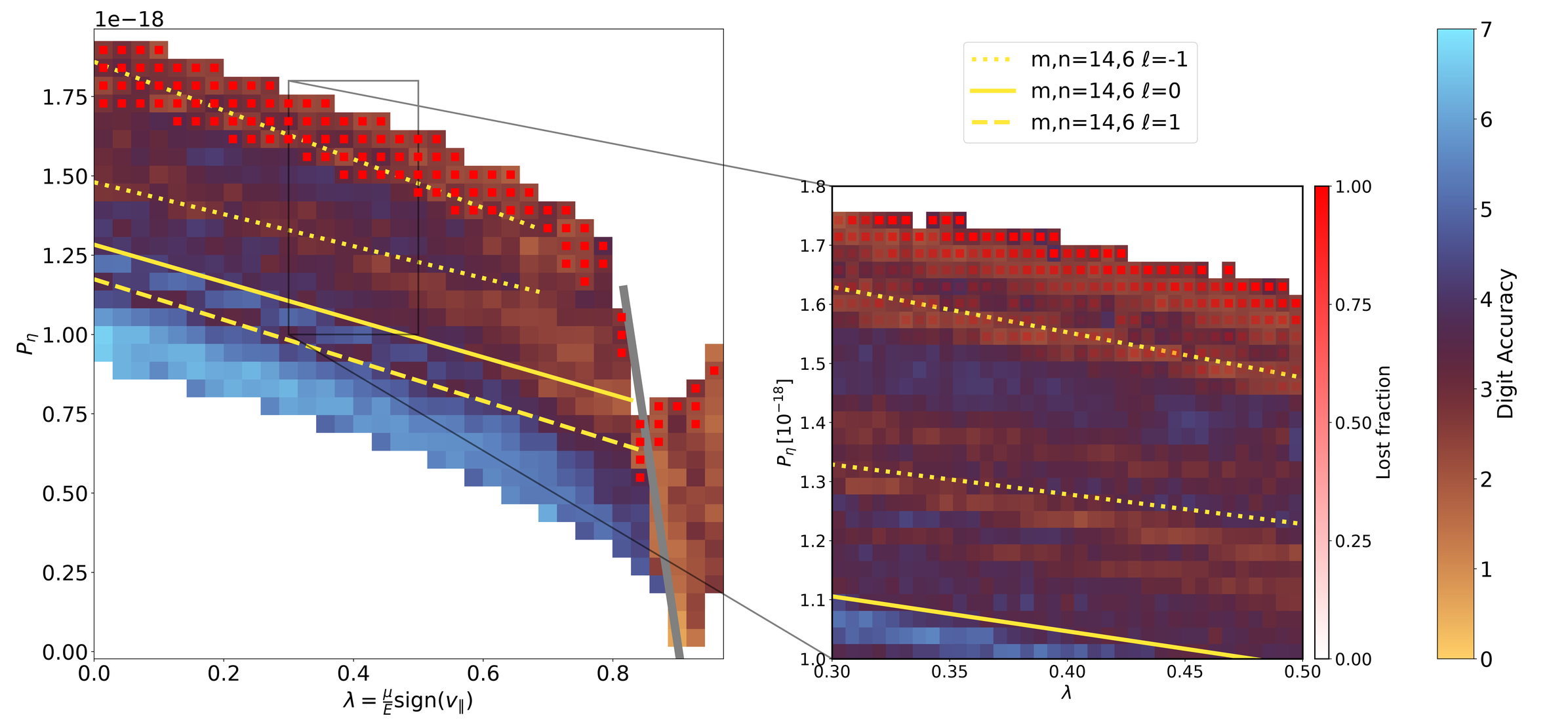}
    
    \caption{Zoomed in segment of the phase-space map demonstrated in Figure \ref{fig:QA4_full} of the LBQA without QS error suppressed. Strata of high and low digit accuracy are visible, together with the losses around the resonance. The zoomed in plot has 35 particles per $(s, \mu)$ bin. In the zoomed in figure, the fraction of particles lost per bin is represented by the transparency of the marker. }
    \label{fig:Zoomed}
    
\end{figure}

Figure \ref{fig:Zoomed} demonstrates a region of high digit accuracy separating regions of low digit accuracy. Particles initialised in the upper region of low digit accuracy are lost. Both the region of low digit accuracy and the region of lost particles correspond with $P_\eta \gtrsim 1.58 \times 10^{-18}$ Js.
The region of higher digit accuracy between $P_\eta \approx 1.45 \times 10^{-18}$ Js and $P_\eta \approx 1.58 \times 10^{-18}$ Js may be acting as a transport barrier, preventing transport between the chaotic layers. 

% Studying AEs without substantial alpha particle losses in the presence of an imperfect background field highlights the effects of omnigenity deviations on EP dynamics. In the following section, we extend this to multiple harmonics to visualize the impact of the whole mode, with and without the QS error.
\section{Phase-space plots for visualising multi-harmonic modes}
\label{sec:multiharmonic}
\subsection{Quasiaxisymmetric equilibrium}
\label{sec:qa_multi}

Having established the phase-space behaviour associated with individual harmonics in both the presence and absence of QS error, we now measure the phase space occupied by chaos due to the full AE. The harmonic truncation follows Section \ref{sec:qs_error}; for the LBQA field this retains 35 harmonics. 
The LBQA experiences substantial alpha-particle losses due to a single harmonic AE compared to the same field with QS-breaking modes suppressed, and also both experience more substantial phase-space chaos in the presence of the full AE. Loss fractions were computed by tracing 10,000 collisionless guiding-centre marker particles for $10^{-2}$ s with FIRM3D, using the fusion-born distribution developed by \cite{bader2021modeling}. In contrast with the single-harmonic perturbations, the superposition of multiple energetic harmonics introduces a larger set of resonances that may overlap, leading to broader regions of stochasticity in phase space. 
We examine the resulting dynamics with and without QS error, at perturbation amplitudes that result in notable particle losses.
\begin{figure}
    \centering
   
        \includegraphics[width=0.95\textwidth]{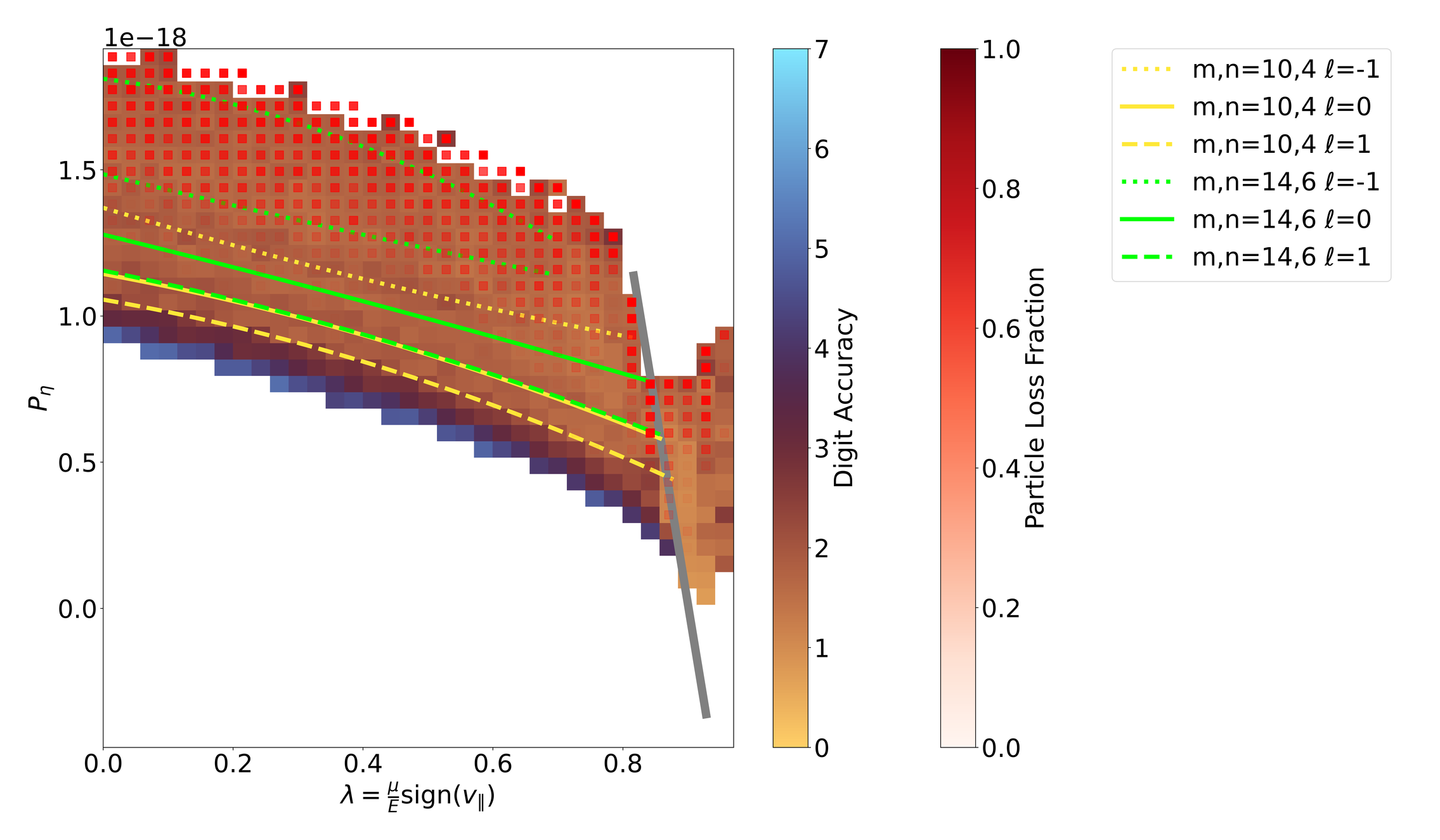}
    
    \caption{Full phase-space map for the LBQA with QS-breaking perturbations. The locations of lost particles have red markers placed upon them. The resonant locations of the two largest-energy harmonics are plotted, but this does not represent all resonances across the 35 harmonics. This plot initialises 30 particles per $(s, \mu)$ bin. In this map, we represent the fraction of particles lost in each bin by the transparency of the red marker. }
    \label{fig:QAFULLAE}
\end{figure}

Figures \ref{fig:QA4_QS} and \ref{fig:QA4_full} isolate the largest-energy harmonic for an AE of perturbation strength $\sqrt{\delta U/U_0}$ = 0.001. The full wave with all harmonics at this perturbation strength corresponds to fusion-born loss fractions of 0.18 in the QS-enforced equilibrium and 0.41 in the equilibrium retaining QS error. In contrast, the fusion-born loss fractions in the single largest-energy harmonic of the AE are
0.003 for the perfect QA, and  0.01 in the equilibrium with the imperfect QA background. Including all 35 harmonics drives increased chaos and losses, demonstrated by both the phase-space maps and the collisionless loss fractions for fusion-born alphas. The phase-space map for the 35-harmonic AE at a strength of $\sqrt{\delta U/U_0}$ = 0.001 is shown in Figure \ref{fig:QAFULLAE}. With and without QS error, the sampled phase space is predominantly stochastic, with regular orbits confined near the core. For this reason, we show only the equilibrium with the QS-breaking perturbations in Figure \ref{fig:QAFULLAE}. 
The overlapping resonances from all 35 harmonics push a large majority of the phase space into stochastic behaviour. 
This may be attributed to the closely spaced resonances in QA, as demonstrated by \eqref{eq:resonance_condition}. Equation \eqref{eq:resonance_condition} can be written in terms of the drift helicity, $h= \left(n-N m-\omega / \omega_\zeta \right)/\left(m+\ell\right)+N$. In QA, $N=0$, so $h=
\left(n-\omega / \omega_\zeta\right)/ \left(m+\ell\right)$, and the resonant location $h$ remains near the primary resonance for each sideband resonance denoted by $\ell$. Closely spaced sideband resonances may explain the large chaotic region demonstrated in the LBQA with the full AE (Figure \ref{fig:QAFULLAE}). 

%The inclusion of static QS breaking perturbations dramatically increases losses. The largest energy resonating harmonic, shown in Figure \ref{fig:QApsmaps}, is dramatically less chaotic than the full harmonic in Figure \ref{fig:QAFULLAE}. This suggests alpha particle loss in the LBQA is primarily driven by resonance island overlap at large amplitudes, and
%Almost the entirety of phase space is chaotic in both the case with and without QS-breaking harmonics, which is radically different than the phase space chaos inflicted by the highest energy resonating harmonic. 
%that island overlap between many harmonics is responsible for EP transport. This behavior could not have been made apparent with Poincaré plots alone. 

\subsection{Quasi-isodynamic equilibrium}
We use phase-space maps to evaluate a QI configuration with the largest-energy harmonic and with the full mode. We use the high-shear SQuID \citep{goodsquid,alex}, with the corresponding 20-harmonic AE. 
Since $P_\eta$ is not conserved in this configuration and no exact constant of the motion is available to label surfaces, we plot phase-space maps by the initial flux coordinate, $s_0$. $P_\eta$ remains the observable that the WBA is applied to, with helicity $M = 0$, $N = N_P$. 
\begin{figure}
    \centering
    \includegraphics[width=\linewidth]{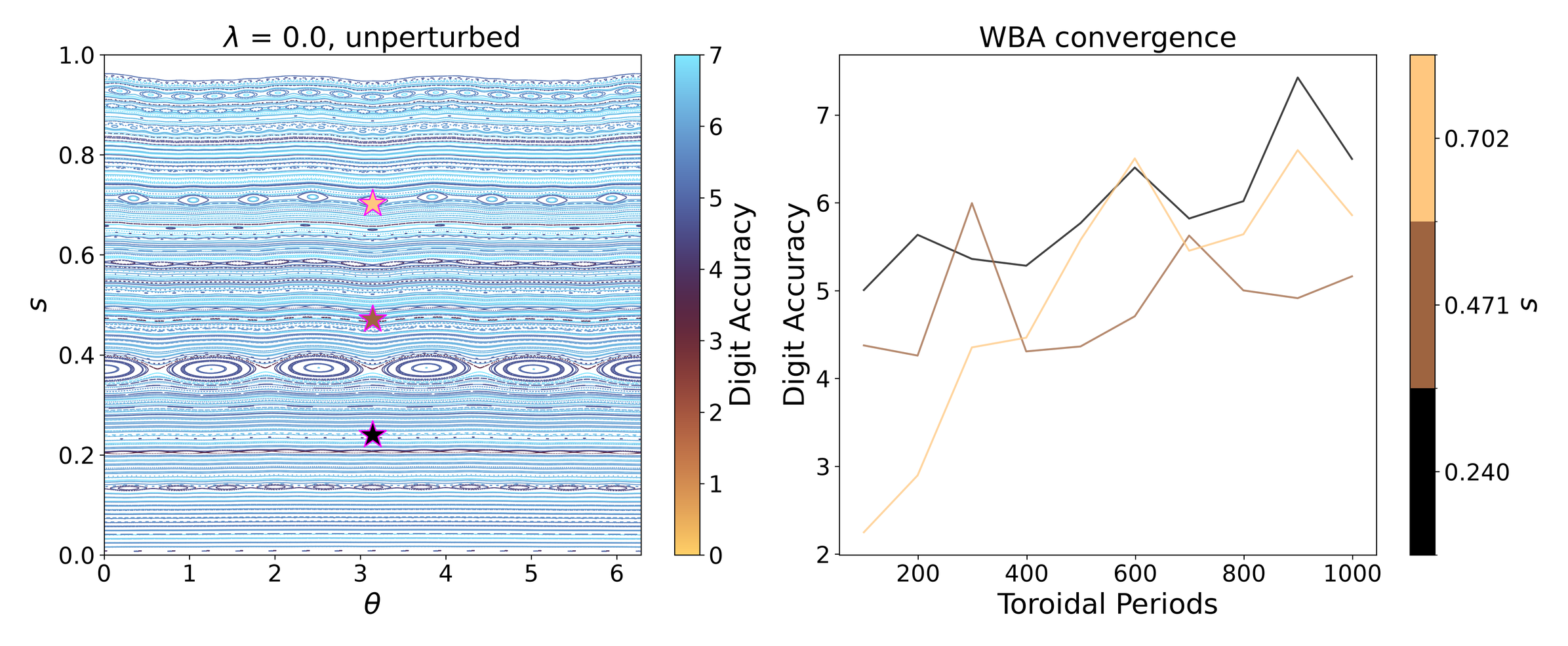}
    \caption{Unperturbed passing Poincaré map for the high-shear SQuID \citep{goodsquid,alex}, along with the corresponding digit accuracy convergence demonstration for selected points. The internal colour of the stars denoted on the Poincaré plot corresponds to the WBA convergence plot. }
    \label{fig:petaproof}
\end{figure}
To ensure that the WBA of $P_\eta$ converges in omnigeneous stellarators in which $P_\eta$ is not necessarily an invariant, we compute the passing map in the equilibrium unperturbed by SAWs (Figure \ref{fig:petaproof}). We find that $P_\eta$ converges as rapidly, and to comparably high digit accuracies as the quasisymmetric cases in this equilibrium.

%\begin{figure}
%    \centering
%    \includegraphics[width=0.6\linewidth]{figures/losses/SQUID_Losses.png}
%    \caption{Fraction of particles lost as a function of $\sqrt{\frac{\delta U}{U_0}}$ for an AE in  the 
%high shear SQuID \citep{goodsquid,alex}. The AE radial structure and harmonics is detailed in Figure \ref{fig:SQUIDAE_struct}.}
%    \label{fig:SQUIDlosses}
%\end{figure}

\begin{figure}
    \centering
    \includegraphics[width=\linewidth]{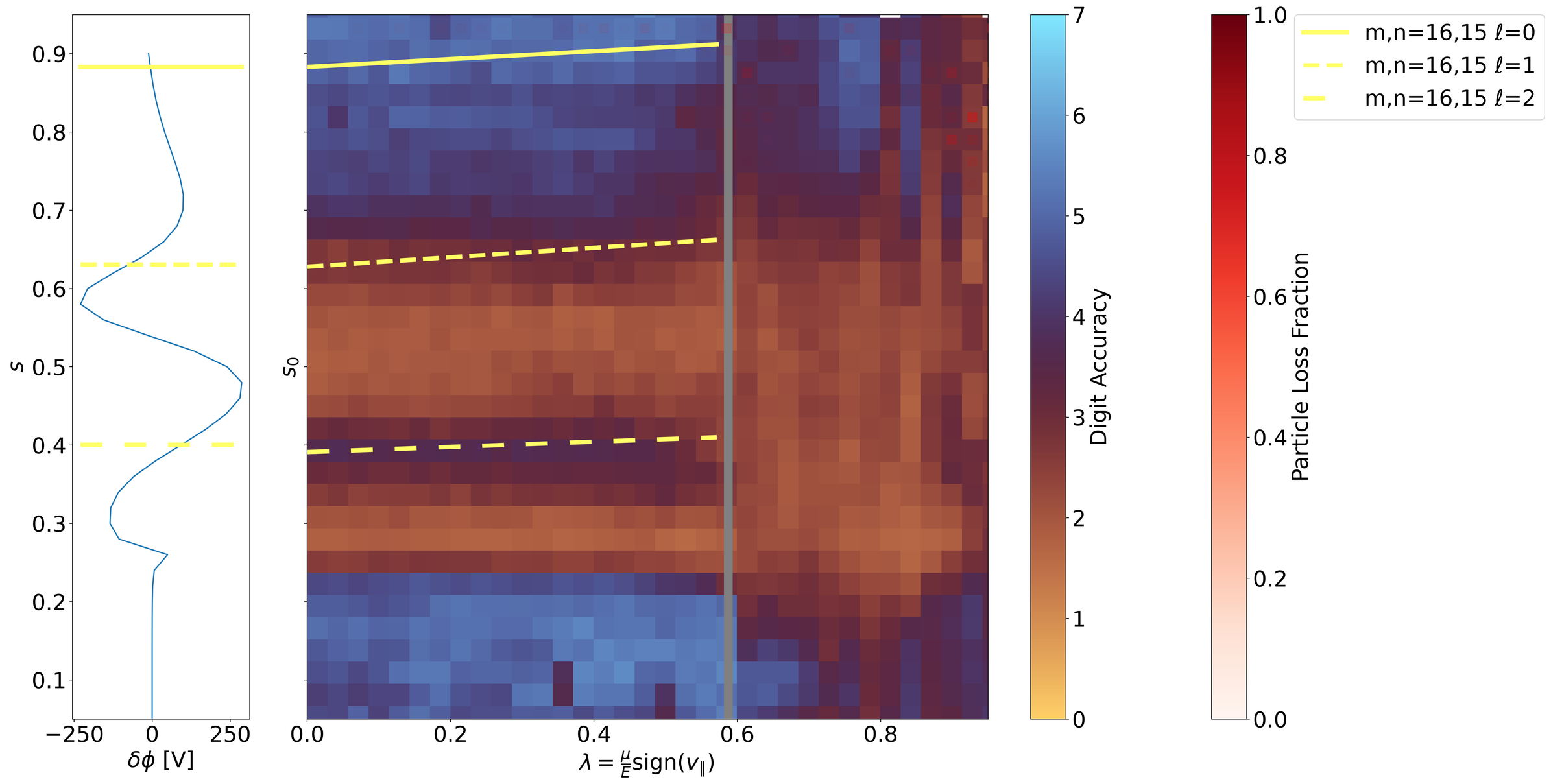}
    \caption{ Phase-space map for the largest-energy $(m, n) = (16, 15)$ harmonic in the 
high-shear SQuID \citep{goodsquid,alex}. This map samples 100 particles per $s_0, \lambda$ bin.}
    \label{fig:SQUIDsingle}
\end{figure}

\begin{figure}
    \centering
    \includegraphics[width=\linewidth]{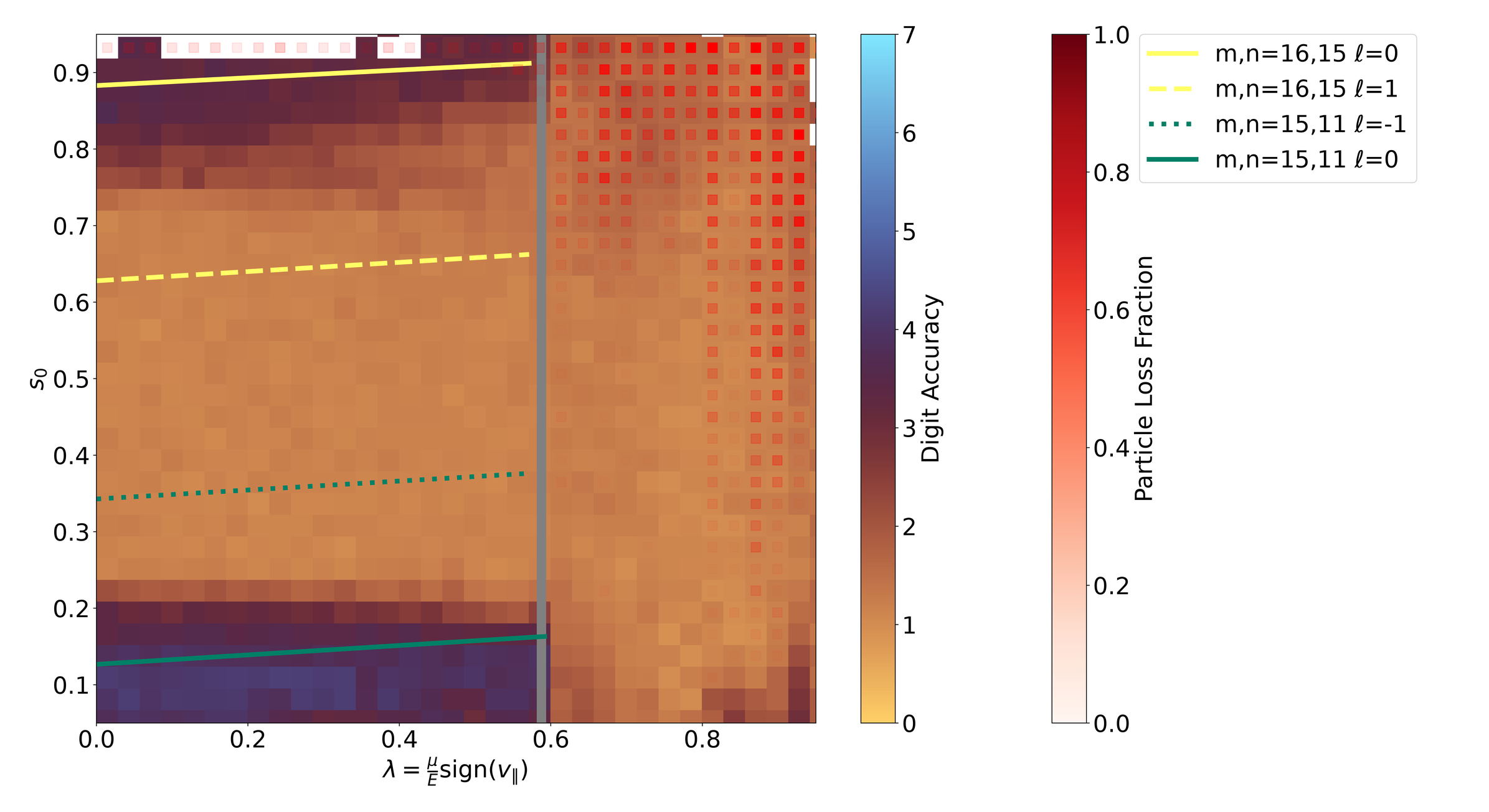}
    \caption{Phase-space map for a 20-harmonic AE in the 
high-shear SQuID \citep{goodsquid,alex}. 
The two largest harmonics have labeled resonance locations, though this does not represent all possible resonances across the 20 harmonics. 
The phase-space plot for the 20-harmonic AE permits visualisation of the trapped region, which becomes entirely chaotic. This map samples 100 particles per $s_0, \lambda$ bin. }
    \label{fig:SQUIDtotal}
\end{figure}

The largest-energy harmonic of the AE is plotted in Figure \ref{fig:SQUIDsingle}, with the full wave in Figure \ref{fig:SQUIDtotal} for a total perturbed energy $\sqrt{\delta U/U_0}$= 0.0021. The sampled phase space of the equilibrium with the full 20 harmonic AE has a chaotic fraction of 0.76, with a fusion-born loss fraction of 0.22 (computed as described in Section \ref{sec:qa_multi}). The complete comparison of chaotic fractions and fusion-born loss fractions is demonstrated in Table \ref{tab:kd}. In the passing region, a band of higher digit accuracy near the edge, beginning at approximately $s_0 = 0.65$, may act as a transport barrier that limits losses from the stochastic region that begins at $s_0 = 0.25$. This may explain the drastic difference between the fraction of chaotic phase space and the fraction of lost particles in this uniformly born distribution of the phase-space map. Additionally, the trapped region exhibits no comparable band of regular orbits. This is consistent with the substantially lower losses observed in the passing region than in the trapped region.
In comparison, the LBQA equilibrium without suppressed QS error, demonstrated in Figure \ref{fig:QAFULLAE}, has a chaotic fraction of 0.80, and a fusion-born loss fraction of 0.41, approximately double that of the SQuID despite the perturbation being half of the energy. Despite having similar chaotic fractions, the high-shear SQuID experiences substantially lower losses, perhaps due to the observed transport barrier. 
% The trapped population of the high shear SQuID exhibits the most significant increase in losses with the full AE. 

Figure \ref{fig:SQUIDsingle} suggests regions of strong harmonic amplitude dominate stochasticity. The stochastic region begins at $s_0=0.25$, below the location of the first resonance, and ends at approximately $s_0=0.62$. This indicates local amplitude has a stronger effect on chaos than resonant surface location.  A region of slightly larger digit accuracy exists at $s_0=0.4$, coinciding with the weaker $\ell = 2$ resonance and approximately where the perturbation crosses $\delta \Phi = 0$.

In Figure \ref{fig:SQUIDtotal}, most particles that are lost originate in the trapped region. Red squares mark $(s_0, \lambda)$ bins in which the appropriate fraction of the 100 particles initialised in that bin was lost. This agrees with loss fractions from the fusion-born distribution reported by \cite{alex}. In several places, indicators of lost particles form almost continuous columns at fixed $\lambda$, extending from a minimum $s_0$ to the wall, suggesting that particles at that $\lambda$ are transported out of the stellarator. Sometimes, these columns are discontinuous, as there may be an individual bin where no particles were lost. These interruptions may occur in regions of phase space where the loss pathway is sensitive to the initial angles $(\theta_0, \zeta_0)$, which determine the magnetic well structure a particle samples along its field line. 
% Sharing the same $s_0$ and $\lambda$ initial condition does not guarantee the same transport qualities for each particle. 
For example, particles may have different trapping states (e.g. ripple versus banana), be trapped in an island or crossing a broken separatrix, among other things. 
For this reason, increased sampling in angles within $s_0$ and $\lambda$ bins may resolve these discontinuities.  
% Bins in $(s_0, \lambda)$ that break the continuity of an intact loss pathway may be chaotic, yet fail to sample the corresponding loss mechanism.  
%Different trapped orbit classes may exhibit different chaotic fractions, and consequently different transport, for example convective or diffusive behaviours \citep{paul_energetic_2022}. In the LBQA, for example, banana trapped particles were generally less chaotic than barely trapped particles (Figure \ref{fig:orbitclassbreakdown}), and a similar class dependence may hold in the high-shear SQuID. 

\begin{table}
  \begin{center}
  \begin{tabular}{lcccc}
       & Harmonics & $\sqrt{\delta U/U_0}$ & Chaotic Fraction & Fusion-born loss fraction \\[10pt] %\hline
      Perfectly QS LBQA  & 1 & $1 \times 10^{-3}$ &  0.007 & 0.003 \\[6pt]
       %& 35  & $1 \times 10^{-3}$ &   & 0.18 \\[6pt]
      LBQA & 1 & $1 \times 10^{-3}$ & 0.36 & 0.01 \\
       & 35 & $1 \times 10^{-3}$ & 0.80 & 0.41 \\[6pt]
    SQuID & 1 & $2 \times 10^{-3}$ & 0.42 & 0.001 \\
        & 20 &  $2 \times 10^{-3}$ & 0.76 & 0.22 \\
  \end{tabular}
  \caption{Comparison of chaotic fractions and Fusion-born loss fraction between the LBQA, the SQuID, and their corresponding SAWs.}
  \label{tab:kd}
  \end{center}
\end{table}
\section{Conclusions}

We have studied two distinct drivers of alpha-particle transport in well-optimised stellarators: deviations from omnigeneity and Alfvén eigenmodes, together with the interaction between them. This work is motivated by the limitations of traditional Poincaré diagnostics in time-dependent perturbed or omnigeneous systems, which rely on assumptions and symmetries to reduce the dynamics to two dimensions \citep{chambliss_fast_2025}.  To address this, we applied weighted Birkhoff averaging (WBA) to the canonical momentum $P_\eta$ along guiding-centre alpha-particle trajectories integrated in FIRM3D \citep{FIRM3DJOSS} as a diagnostic for the onset of chaos. As WBA requires only a smooth observable and not an exact invariant of the motion, the diagnostic remains well posed where the Poincaré map does not. We show that the convergence of the average of $P_\eta$ continues to diagnose orbit regularity even where $P_\eta$ is not itself conserved, for example in a general omnigeneous stellarator.

This work introduces a WBA phase-space map, which represents the digit accuracy of particles initialised across alpha-particle phase space parameterized by a radial and pitch angle coordinate. To complete our orbit classifications, we first calibrated a chaos threshold digit accuracy of 3, defined in \eqref{eq:DA}, by direct comparison against structures resolved in a Poincaré map and by comparing the distribution of the digit accuracy of alpha particles lost due to a SAW, excluding equilibrium losses to ensure the losses are attributable to broken invariant tori. We find that the fraction of particles classified as chaotic is insensitive to tightening the guiding-centre integration tolerance, saturating beyond a tolerance of $10^{-8}$ (Figure \ref{fig:integration}).

We applied this framework to configurations close to omnigeneity, with and without SAWs. For the LBQA stellarator without a SAW, we find that the onset of chaos occurs at the passing/trapped boundary, and that most confined trapped particles are chaotic. Additionally, in passing particles, deviations from precise quasisymmetry of the equilibrium produce resonant islands, with chaos stemming from both large islands experiencing separatrix breaking, and overlapping resonances. We then enforce perfect quasisymmetry, and add a SAW harmonic. In the LBQA without enforced quasisymmetry and a single SAW harmonic, we find that the chaotic regions are significantly broadened compared with the perfectly quasisymmetric case, and previously intact islands now exhibit chaotic behaviour. We find the extent of chaotic layers is governed by the local amplitude of the wave, and is less dependent on the location of the resonant surface. Additionally, we find that single-harmonic perturbations, corresponding to a single $(m, n, \omega)$ set, are not responsible for most chaos in phase space in both the LBQA \citep{landreman2022optimization} and the high-shear SQuID \citep{goodsquid,alex}. Although the inclusion of a single SAW harmonic causes minimal phase-space chaos in a perfectly quasisymmetric equilibrium at sufficiently small amplitudes, chaotic regions of phase space grow due to the interaction between harmonics and omnigeneity deviations, demonstrated in both the LBQA and the high-shear SQuID. 

Several limitations bound the use of this method, such as the use of collisionless tracing, and the marker density required for statistically meaningful bin averages. WBA would not be applicable to collisional processes beyond drag, as stochastic collisional processes such as pitch-angle scattering destroy the quasiperiodic motion on which WBA relies. 
% In this work, the digit accuracy is evaluated at $T = 10^{-2}$ s, approximately one tenth of the alpha-particle slowing-down time. 
Additionally, the most computationally expensive part of this work is the guiding-centre particle tracing, which is exacerbated by the density of sampled particles needed to populate the map. Dense sampling of $(P_\eta, \mu)$ or $(s, \mu)$ is necessary, since, for example, one bin in $(s, \mu)$ may sample both ripple and banana trapped classes depending on the initial angles $(\theta_0, \zeta_0)$, with different corresponding locations in magnetic wells. For this reason, this work may benefit from methods such as that of \cite{MRuth}, which adapts both the averaging weights and the integration time to each trajectory and may reduce the tracing cost, though it has so far been demonstrated only for two-dimensional maps.

In future work we intend to connect this picture to the nonlinear evolution of Alfvén eigenmodes. 
% As established in this work, when the field is not perfectly omnigenous, resonant islands in drift space can become stochastic. 
Nonlinear AE behaviour, such as frequency chirping, depends on the persistence of coherent phase-space structures.
% within these islands.
% , which can be made stochastic by imperfections in omnigeneity. 
Previous work has shown that sufficiently stochastic behaviour, such as diffusion induced by microturbulence, can prevent the necessary resonant coherence with the wave for nonlinear modulation of frequency in AEs \citep{duarte_theory_2017}. Imperfections in omnigeneity may therefore play a role analogous to microturbulence in setting the nonlinear evolution of AEs.
% Future work hopes to establish the relationship between AE resonant islands made stochastic by imperfect omnigeneity and the onset of nonlinear AE behaviour.

\bibliographystyle{jpp}

\bibliography{jpp-instructions}

@article{landreman2026bayesian,
  title={Bayesian optimization of stellarator alpha-particle confinement using data-informed parameter spaces and dimensionality reduction},
  author={M. Landreman and M. Czekanski and A. Giuliani and B. Jang and R. Conlin},
  journal={arXiv preprint arXiv:2606.19523},
  year={2026}
}

@article{kolmogorov1954,
  author  = {Kolmogorov, A. N.},
  title   = {On the conservation of conditionally periodic motions
             for a small change in {H}amilton's function},
  journal = {Doklady Akademii Nauk SSSR},
  volume  = {98},
  pages   = {527--530},
  year    = {1954}
}

@article{arnold1963proof,
  author  = {Arnold, V. I.},
  title   = {Proof of a theorem of {A}. {N}. {K}olmogorov on the invariance
             of quasi-periodic motions under small perturbations
             of the {H}amiltonian},
  journal = {Russian Mathematical Surveys},
  volume  = {18},
  number  = {5},
  pages   = {9--36},
  year    = {1963}
}

@article{moser1962,
  author  = {Moser, J.},
  title   = {On invariant curves of area-preserving mappings of an annulus},
  journal = {Nachrichten der Akademie der Wissenschaften in G\"ottingen,
             II. Mathematisch-Physikalische Klasse},
  pages   = {1--20},
  year    = {1962}
}

@article{meissdun,
  author    = {N. Duignan and J. Meiss},
  title     = {Distinguishing between regular and chaotic orbits of flows by the weighted {B}irkhoff average},
  journal   = {Physica D: Nonlinear Phenomena},
  volume    = {449},
  pages     = {133749},
  year      = {2023},
  doi       = {10.1016/j.physd.2023.133749},
}

@article{quasisym_landreman_paul,
  author    = {M. Landreman and E. J. Paul},
  title     = {Magnetic fields with precise quasisymmetry for plasma confinement},
  journal   = {Physical Review Letters},
  volume    = {128},
  number    = {3},
  pages     = {035001},
  year      = {2022},
  publisher = {APS},
}

@article{transportstellvtokamak,
  author    = {K. Toi and K. Ogawa and M. Isobe and M. Osakabe and D. A. Spong and Y. Todo},
  title     = {Energetic-ion-driven global instabilities in stellarator/helical plasmas and comparison with tokamak plasmas},
  journal   = {Plasma Physics and Controlled Fusion},
  volume    = {53},
  number    = {2},
  pages     = {024008},
  year      = {2011},
  publisher = {IOP Publishing},
}

@article{Albert_alpha_transport,
  author    = {C. G. Albert and K. Rath and R. Babin and R. Buchholz and S. V. Kasilov and W. Kernbichler},
  title     = {Resonant transport of fusion alpha particles in quasisymmetric stellarators},
  journal   = {Journal of Physics: Conference Series},
  volume    = {2397},
  number    = {1},
  pages     = {012009},
  year      = {2022},
  publisher = {IOP Publishing},
  doi       = {10.1088/1742-6596/2397/1/012009},
}

@article{ITER,
  author    = {N. N. Gorelenkov and S. D. Pinches and K. Toi},
  title     = {Energetic particle physics in fusion research in preparation for burning plasma experiments},
  journal   = {Nuclear Fusion},
  volume    = {54},
  number    = {12},
  pages     = {125001},
  year      = {2014},
  month     = nov,
  publisher = {IOP Publishing},
  doi       = {10.1088/0029-5515/54/12/125001},
}

@article{Bindel_2023,
  author    = {D. Bindel and M. Landreman and M. Padidar},
  title     = {Direct Optimization of Fast-Ion Confinement in Stellarators},
  journal   = {Plasma Physics and Controlled Fusion},
  volume    = {65},
  number    = {6},
  pages     = {065012},
  year      = {2023},
  month     = may,
  publisher = {IOP Publishing},
  doi       = {10.1088/1361-6587/acd141},
}

@article{ASDEX,
  author    = {J. Gonzalez-Martin and others},
  note      = {{ASDEX Upgrade Team and EUROfusion MST1 Team}},
  title     = {Active Control of {Alfvén} Eigenmodes by Externally Applied {3D} Magnetic Perturbations},
  journal   = {Physical Review Letters},
  volume    = {130},
  number    = {3},
  pages     = {035101},
  year      = {2023},
  month     = jan,
  doi       = {10.1103/PhysRevLett.130.035101},
}

@article{NSTX,
  author    = {A. Bortolon and W. W. Heidbrink and G. J. Kramer and J.-K. Park and E. D. Fredrickson and J. D. Lore and M. Podestà},
  title     = {Mitigation of {Alfvén} Activity in a Tokamak by Externally Applied Static {3D} Fields},
  journal   = {Physical Review Letters},
  volume    = {110},
  number    = {26},
  pages     = {265008},
  year      = {2013},
  month     = jun,
  doi       = {10.1103/PhysRevLett.110.265008},
}

@article{Boozer1983,
  author    = {A. H. Boozer},
  title     = {Transport and isomorphic equilibria},
  journal   = {Physics of Fluids},
  volume    = {26},
  number    = {2},
  pages     = {496--503},
  year      = {1983},
  publisher = {AIP Publishing},
  doi       = {10.1063/1.864166},
}

@misc{FIRM3DJOSS,
  author        = {E. Paul and A. Knyazev and M. Czekanski and A. Lachmann and A. Hyder and C. Albert and M. Landreman},
  title         = {{FIRM3D}: Fast ion reduced models in {3D}},
  year          = {2026},
  eprint        = {2605.16734},
  archiveprefix = {arXiv},
  primaryclass  = {physics.plasm-ph},
  url           = {https://arxiv.org/abs/2605.16734},
}

@article{landreman2022optimization,
  author    = {M. Landreman and S. Buller and M. Drevlak},
  title     = {Optimization of quasi-symmetric stellarators with self-consistent bootstrap current and energetic particle confinement},
  journal   = {Physics of Plasmas},
  volume    = {29},
  number    = {8},
  pages     = {082501},
  year      = {2022},
  publisher = {AIP Publishing},
}

@article{Mynicklown,
  author    = {H. E. Mynick},
  title     = {Transport of energetic ions by low-n magnetic perturbations},
  journal   = {Physics of Fluids B: Plasma Physics},
  volume    = {5},
  number    = {5},
  pages     = {1471--1481},
  year      = {1993},
  month     = may,
  doi       = {10.1063/1.860886},
}

@article{paul_energetic_2022,
  author    = {E. J. Paul and A. Bhattacharjee and M. Landreman and D. Alex and J. L. Velasco and R. Nies},
  title     = {Energetic particle loss mechanisms in reactor-scale equilibria close to quasisymmetry},
  journal   = {Nuclear Fusion},
  volume    = {62},
  number    = {12},
  pages     = {126054},
  year      = {2022},
  month     = dec,
  doi       = {10.1088/1741-4326/ac9b07},
}

@misc{alex,
  author        = {A. Knyazev and A. Lachmann and A. Goodman and A. Hyder and M. Czekanski and D. Spong and E. Paul},
  title         = {On shear {Alfvén} wave-induced energetic ion transport in optimized stellarators},
  year          = {2026},
  eprint        = {2603.03118},
  archiveprefix = {arXiv},
  primaryclass  = {physics.plasm-ph},
  url           = {https://arxiv.org/abs/2603.03118},
}

@article{spongstability,
  author    = {D. A. Spong and E. D'azevedo and Y. Todo},
  title     = {Clustered frequency analysis of shear {Alfvén} modes in stellarators},
  journal   = {Physics of Plasmas},
  volume    = {17},
  number    = {2},
  pages     = {022106},
  year      = {2010},
  publisher = {AIP Publishing},
}

@article{Paul_Mynick_Bhattacharjee_2023,
  author    = {E. J. Paul and H. E. Mynick and A. Bhattacharjee},
  title     = {Fast-ion transport in quasisymmetric equilibria in the presence of a resonant {Alfvénic} perturbation},
  journal   = {Journal of Plasma Physics},
  volume    = {89},
  number    = {5},
  pages     = {905890515},
  year      = {2023},
  doi       = {10.1017/S0022377823001095},
}

@article{littlejohn_variational_1983,
  author    = {R. G. Littlejohn},
  title     = {Variational principles of guiding centre motion},
  journal   = {Journal of Plasma Physics},
  volume    = {29},
  number    = {1},
  pages     = {111--125},
  year      = {1983},
  month     = feb,
  doi       = {10.1017/S002237780000060X},
}

@article{rotationmetric,
  author    = {N. Kallinikos and R. S. MacKay and D. Martínez-del-Río},
  title     = {Regions without flux surfaces of given class for magnetic fields in toroidal geometry},
  journal   = {Plasma Physics and Controlled Fusion},
  volume    = {65},
  number    = {9},
  pages     = {095021},
  year      = {2023},
  month     = sep,
  doi       = {10.1088/1361-6587/acea3f},
}

@article{White_2011,
  author    = {R. B. White},
  title     = {Modification of particle distributions by magnetohydrodynamic instabilities {II}},
  journal   = {Plasma Physics and Controlled Fusion},
  volume    = {53},
  number    = {8},
  pages     = {085018},
  year      = {2011},
  month     = jun,
  doi       = {10.1088/0741-3335/53/8/085018},
}

@article{bader2021modeling,
  author    = {A. Bader and D. T. Anderson and M. Drevlak and B. J. Faber and C. C. Hegna and S. Henneberg and M. Landreman and J. C. Schmitt and Y. Suzuki and A. Ware},
  title     = {Modeling of energetic particle transport in optimized stellarators},
  journal   = {Nuclear Fusion},
  volume    = {61},
  number    = {11},
  pages     = {116060},
  year      = {2021},
  publisher = {IOP Publishing},
}

@article{goodsquid,
  author    = {A. G. Goodman and K. Camacho Mata and S. A. Henneberg and R. Jorge and M. Landreman and G. G. Plunk and H. M. Smith and R. J. J. Mackenbach and P. Helander},
  title     = {Constructing precisely quasi-isodynamic magnetic fields},
  journal   = {Journal of Plasma Physics},
  volume    = {89},
  number    = {5},
  pages     = {905890504},
  year      = {2023},
  doi       = {10.1017/S002237782300065X},
}

@article{duarte_theory_2017,
  author    = {V. N. Duarte and H. L. Berk and N. N. Gorelenkov and W. W. Heidbrink and G. J. Kramer and R. Nazikian and D. C. Pace and M. Podestà and M. A. Van Zeeland},
  title     = {Theory and observation of the onset of nonlinear structures due to eigenmode destabilization by fast ions in tokamaks},
  journal   = {Physics of Plasmas},
  volume    = {24},
  number    = {12},
  pages     = {122508},
  year      = {2017},
  month     = dec,
  doi       = {10.1063/1.5007811},
}

@article{Infinity2EPs,
  author    = {L. Carbajal and J. Varela and others},
  title     = {Alpha-particle confinement in {Infinity Two} Fusion Pilot Plant baseline plasma design},
  journal   = {Journal of Plasma Physics},
  volume    = {91},
  number    = {4},
  pages     = {E94},
  year      = {2025},
  month     = aug,
  doi       = {10.1017/S0022377825000352},
}

@article{SPARC_Physics,
  author    = {P. Rodriguez-Fernandez and A. J. Creely and others},
  title     = {Overview of the {SPARC} physics basis towards the exploration of burning-plasma regimes in high-field, compact tokamaks},
  journal   = {Nuclear Fusion},
  volume    = {62},
  number    = {4},
  pages     = {042003},
  year      = {2022},
  month     = sep,
  doi       = {10.1088/1741-4326/ac1654},
}

@article{chambliss_fast_2025,
  author    = {A. Chambliss and E. Paul and S. R. Hudson},
  title     = {Fast particle trajectories and integrability in quasiaxisymmetric and quasihelical stellarators},
  journal   = {Journal of Plasma Physics},
  volume    = {91},
  number    = {3},
  pages     = {E74},
  year      = {2025},
  doi       = {10.1017/S0022377825000431},
}

@article{SPARC_EPs,
  author    = {S. D. Scott and G. J. Kramer and E. A. Tolman and A. Snicker and J. Varje and K. Särkimäki and J. C. Wright and P. Rodriguez-Fernandez},
  title     = {Fast-ion physics in {SPARC}},
  journal   = {Journal of Plasma Physics},
  volume    = {86},
  number    = {5},
  pages     = {865860508},
  year      = {2020},
  month     = oct,
  doi       = {10.1017/S0022377820001087},
}

@article{MRuth,
    author = {Ruth, M. and Bindel, D.},
    title = {Finding {B}irkhoff averages via adaptive filtering},
    journal = {Chaos: An Interdisciplinary Journal of Nonlinear Science},
    volume = {34},
    number = {12},
    pages = {123109},
    year = {2024},
    month = {12},
    issn = {1054-1500},
    doi = {10.1063/5.0215396},
    url = {https://doi.org/10.1063/5.0215396}
}

@misc{Foster,
      title={Energetic-particle orbits near rational flux surfaces in stellarators: {I}. {P}assing particles}, 
      author={T. E. Foster and F. I. Parra and R. B. White and J. L. Velasco and I. Calvo and E. J. Paul},
      year={2025},
      eprint={2512.03165},
      archivePrefix={arXiv},
      primaryClass={physics.plasm-ph},
      url={https://arxiv.org/abs/2512.03165}, 
}

@article{BeidlerTransitioning,
    author = {Beidler, C. D. and Kolesnichenko, Ya. I. and Marchenko, V. S. and Sidorenko, I. N. and Wobig, H.},
    title = {Stochastic diffusion of energetic ions in optimized stellarators},
    journal = {Physics of Plasmas},
    volume = {8},
    number = {6},
    pages = {2731-2738},
    year = {2001},
    month = {06},
    issn = {1070-664X},
    doi = {10.1063/1.1365958},
    url = {https://doi.org/10.1063/1.1365958},
}

@article{heidbrink2008basic,
  author    = {W. W. Heidbrink},
  title     = {Basic physics of {Alfvén} instabilities driven by energetic particles in toroidally confined plasmas},
  journal   = {Physics of Plasmas},
  volume    = {15},
  number    = {5},
  pages     = {055501},
  year      = {2008},
  doi       = {10.1063/1.2838239},
}

@article{chirikov1979,
  author    = {B. V. Chirikov},
  title     = {A universal instability of many-dimensional oscillator systems},
  journal   = {Physics Reports},
  volume    = {52},
  number    = {5},
  pages     = {263--379},
  year      = {1979},
  doi       = {10.1016/0370-1573(79)90023-1},
}

@article{sander2020birkhoff,
  author    = {E. Sander and J. D. Meiss},
  title     = {{B}irkhoff averages and rotational invariant circles for area-preserving maps},
  journal   = {Physica D: Nonlinear Phenomena},
  volume    = {411},
  pages     = {132569},
  year      = {2020},
  doi       = {10.1016/j.physd.2020.132569},
}

\end{document}